%% file: main.tex
\documentclass[fleqn,usenatbib]{mnras}
\usepackage{newtxtext,newtxmath}
\usepackage[T1]{fontenc}

\DeclareRobustCommand{\VAN}[3]{#2}
\let\VANthebibliography\thebibliography
\def\thebibliography{\DeclareRobustCommand{\VAN}[3]{##3}\VANthebibliography}

\usepackage{graphicx}	
\usepackage{amsmath}	

\usepackage[normalem]{ulem} 
\usepackage[dvipsnames]{xcolor} 
\usepackage{wasysym} 
\definecolor{darkgreen}{rgb}{0.0,0.55,0.0}
\definecolor{darkblue}{rgb}{0.0,0.0,0.5}

\usepackage{soul}
\newcommand{\eg}[0]{$\textnormal{e.g. }$}
\newcommand{\ie}[0]{$\textnormal{i.e. }$}
\newcommand{\Msun}[0]{\,\textnormal{M}_{\textnormal{\astrosun}}}

\newcommand{\tn}[1]{\textnormal{#1}}
\newcommand{\sub}[1]{_{\textnormal{#1}}}

\newcommand{\lgal}[0]{\textsc{L-Galaxies}}
\newcommand{\lgaltt}[0]{\textsc{L-Galaxies 2020}}
\newcommand{\mil}[0]{\textsc{Millennium}}

\newcommand{\Zhalo}[0]{[Fe/H]$_{*,\tn{halo}}$}
\newcommand{\FeH}[0]{$\langle$[Fe/H]$\rangle$}
\newcommand{\MZhR}[0]{M$\tn{Z}\sub{h}$R}

\usepackage{newtxtext,newtxmath}

\title[Stellar haloes in L-GALAXIES 2020]{L-GALAXIES 2020: The formation and chemical evolution of stellar haloes in Milky Way Analogues and galaxy clusters}

\author[Geoff G. Murphy, Robert M. Yates, Shazrene S. Mohamed]{
Geoff G. Murphy,$^{1,2}$\thanks{E-mail: geoffmurphy@saao.ac.za}
Robert M. Yates,$^{3}$
Shazrene S. Mohamed$^{1,2,4}$
\\
$^{1}$ South African Astronomical Observatory, P.O Box 9, Observatory, 7935, Cape Town, South Africa\\
$^{2}$Department of Astronomy, University of Cape Town, Private Bag X3, Rondebosch 7701, South Africa\\
$^{3}$Astrophysics Research Group, University of Surrey, Stag Hill, University Campus, Guildford GU2 7XH, United Kingdom\\
$^{4}$National Institute for Theoretical Physics (NITheP), KwaZulu-Natal, South Africa
}

\date{Accepted XXX. Received YYY; in original form ZZZ}

\pubyear{2021}

\begin{document}
\label{firstpage}
\pagerange{\pageref{firstpage}--\pageref{lastpage}}
\maketitle

\begin{abstract}
We present an analysis of the formation and chemical evolution of stellar haloes around (a) Milky Way Analogue (MWA) galaxies and (b) galaxy clusters in the \lgaltt{} semi-analytic model of galaxy evolution. Observed stellar halo properties are better reproduced when assuming a gradual stripping model for the removal of cold gas and stars from satellites, compared to an instantaneous stripping model. The slope of the stellar mass -- metallicity relation for MWA stellar haloes is in good agreement with that observed in the local Universe. This extends the good agreement between \lgaltt{} and metallicity observations from the gas and stars inside galaxies to those outside. Halo stars contribute on average only $\sim{}0.1$ per cent of the total circumgalactic medium (CGM) enrichment by $z=0$ in MWAs, ejecting predominantly carbon produced by AGB stars. Around a quarter of MWAs have a single `significant progenitor' with a mean mass of $\sim$\,$2.3\times{}10^{9}\Msun$, similar to that measured for Gaia Enceladus. For galaxy clusters, \lgaltt{} shows good correspondence with observations of stellar halo mass fractions, but slightly over-predicts stellar masses. Assuming a Navarro-Frenk-White profile for the stellar halo mass distribution provides the best agreement. On average, the intracluster stellar component (ICS) is responsible for 5.4 per cent of the total intracluster medium (ICM) iron enrichment, exceeding the contribution from the brightest cluster galaxy (BCG) by $z=0$. We find that considering gradual stripping of satellites and realistic radial profiles is crucial for accurately modelling stellar halo formation on all scales in semi-analytic models.
\end{abstract}


\section{Introduction}
\label{sec:Intro}

\input{Chapters/Introduction}

\section{L-GALAXIES 2020} 
\label{sec:Simulation}
In this section, we present an overview of the \lgaltt{} model, including the prescriptions of most relevance for stellar halo studies. We also present the analysis methods applied to the model data, which allow for fair comparisons to literature results.
\input{Chapters/Simulation}

\input{Chapters/Analysis}

\section{Results}
\label{sec:Results}
In this section, we present stellar halo iron abundance results, interaction histories, and hot gas enrichment histories on galactic scales. On galaxy cluster scales, ICS masses and mass fraction results are presented.

\subsection{Milky Way analogue (MWA) galaxies}
\label{sec:GalaxyResults}

\input{Chapters/Results/GalaxyResults}

\subsection{Galaxy clusters}
\label{sec:GalaxyClusters}
\input{Chapters/Results/GalaxyClusterResults}

\section{Conclusions}
\label{sec:Conclusions}
\input{Chapters/Conclusions}

\section*{Acknowledgements}

We thank the referee for their insightful comments and suggestions. We also thank Daniel Cunnama for his input on the MSc project this work is based on, as well as Payel Das and Bruno Henriques for their helpful feedback. G.M. and S.M. acknowledge funding from the South African National Research Foundation and University of Cape Town VC2030 Future Leaders Award (PI: S.M.). The simulations were run on the SAAO computing cluster. Analysis made significant use of \textsc{python} 3.7.4, and the associated packages \textsc{numpy}, \textsc{matplotlib}, \textsc{pandas}, \textsc{seaborn}, and \textsc{scipy}.

\section*{Data Availability}
\lgaltt{} source code is made publicly available at \href{https://lgalaxiespublicrelease.github.io/}{lgalaxiespublicrelease.github.io/}, as are example output catalogues. Complete output catalogues from the default model are also available via the \textsc{Millennium} database at \href{http://gavo.mpa-garching.mpg.de/Millennium/}{gavo.mpa-garching.mpg.de/Millennium/}. The additional \lgal{} data presented here can be obtained from the corresponding author upon request.



\bibliographystyle{mnras}
\bibliography{references,robyates} 







\bsp	
\label{lastpage}
\end{document}

%% file: Chapters/Introduction.tex
A diffuse stellar component permeating the space between galaxies was first discovered in the Coma cluster by \citet{Zwicky_1951}, who described it as "vast and often very irregular swarms of stars...between the conventional spiral, elliptical, and irregular galaxies". Today, this component is referred to as intracluster stars (ICS). Evidence of a similarly diffuse component surrounding our own galaxy was observed in later years, for example by \citet{1962ApJ...136..748E}. This has come to be known as the stellar halo. It was suspected that the stellar halo contained imprints of the formation history of the Milky Way, and attempts were made to understand its formation (\eg{}\citealt{1978ApJ...225..357S}). Dissipative and dissipationless formation mechanisms were put forth, which correspond to ``in-situ'' and ``ex-situ'' formation, respectively \citep{2019ApJ...887..237C}. The advent of CCDs, however, made the study of the dim stellar halo a great deal more feasible, and it was found that haloes are commonplace components of galaxies undergoing continuous evolution \citep{2019arXiv191201616M}. A detailed discussion on the history and current knowledge of intracluster light (ICL), and by extension stellar haloes, is provided by \citet{2019arXiv191201616M}, as well as the challenges associated with its observation.

Stellar haloes have increasingly become a subject of study as they contain important clues and information about the interaction histories of galaxies and galaxy clusters. \citet{2018Natur.563...85H} and \citet{2018MNRAS.478..611B}, for example, found evidence of an accretion event of a satellite galaxy in the Milky Way's inner stellar halo, known as the Gaia Enceladus stream. Stellar streams such as this one are one of the clearest indicators of past interactions with satellite galaxies in the Milky Way stellar halo (\eg{}\citealt{Belokurov+06}). \citet{Ibata+19} identified eight high-significance streams in the inner stellar halo from \textit{Gaia} Data Release 2 (DR2) data \citep{2018A&A...616A...1G}. Similarly, \citet{2019MNRAS.488.1235M} identified a stream attributed to the Sequoia Event, the remnant of the Sequoia dwarf galaxy accretion in the Milky Way's inner halo, and \citet{2020A&A...638A.104R} obtained position, velocity, and metallicity measurements to accurately determine the extent of the Sagittarius stream.

Satellite galaxies which have helped form the Milky Way's stellar halo also leave evidence in the form of globular clusters (GC). For example, \citet{2010A&A...511L..10N} studied the alpha enhancement ([$\alpha$/Fe]) of stellar halo components in the solar neighbourhood. They suggest that low-alpha-enhanced stars could have come from a progenitor dwarf galaxy whose nucleus is now the $\omega$ Cen GC. Another example is the recently discovered LMS-1 stream, which is linked to two GCs \citep{2020ApJ...898L..37Y}. \citet{2021A&A...650L..11M} discovered a GC which they find to be associated with the Hrid halo stream. This stream was one of two new detections by \citet{2021ApJ...914..123I}, along with seven known streams, using the \textit{Gaia} DR2 and eDR3 \citep{2021A&A...649A...1G}.

Observations also suggest that the Milky Way's inner halo is comparatively more metal rich than its outer halo. As such, the stellar halo is often considered to consist of two components: a metal-richer inner halo, and a metal-poorer outer halo. The former is comprised of accreted stars as well as stars dissipated from the disc and bulge of the Milky Way, while the latter is formed mainly from the stellar stripping of dwarf satellite galaxies. \citep{2010A&A...511L..10N,2012A&A...538A..21S,2013ApJ...763...65A,2015ApJ...813L..28A}. 

As a result of these varying formation mechanisms, these two components show a number of differing properties apart from their relative metallicities. For example, \citet{2010A&A...511L..10N} found that the inner and outer haloes show a separation in [$\alpha$/Fe]. Also, using stars from the SDSS catalogue, \citet{2007Natur.450.1020C,2010ApJ...712..692C} are two of the initial works to definitively show that the Milky Way's stellar halo could be partitioned into two components based on density profiles, orbits, and metallicities. In those studies, the inner halo was found to display small to no prograde motion, on average, while the outer halo showed an overall retrograde motion.

In light of the above discussion, a complimentary way of classifying stellar halo stars is according to whether they are of "ex-situ" or "in-situ" origin. The former is dominant at larger radii and includes stars born in satellite galaxies and later stripped into the stellar halo. The latter is dominant at smaller radii and can include stars scattered out of the Milky Way's thick disc, formed from halo gas pre-stripped from satellites, or formed from gas smoothly accreted in the first stages of assembly (see \citealt{Cooper+15}). 

There are now an increasing number of Milky Way stellar halo studies, with investigations ranging from density and metallicity profiles \citep{2015ApJ...809..144X,2016MNRAS.460.1725D,2017ApJ...841...59Z,2019ApJ...887..237C}, to measurements of the total stellar mass \citep{2008ApJ...680..295B,2011MNRAS.416.2903D,2019MNRAS.490.3426D}. Similar studies extend to the stellar haloes of nearby galaxies, for example \citet{2016ApJ...830...62M} and \citet{2016MNRAS.457.1419M}.

On galaxy cluster scales, the ICS contains non-negligible mass, and is a key component in understanding the properties of these large-scale structures which are thought to have a baryon content representative of the Universe's overall baryon fraction \citep{2007ApJ...666..147G}. Accurate total cluster mass measurements, therefore, need to include the ICS. As with single galaxies, the ICS in galaxy clusters contains information about galaxy interaction histories. Studies which measure ICL include \citet{2011A&A...533A.138C} and \citet{2019MNRAS.482.2838M}, while \citet{2011ApJ...729..142S} infer the properties of cluster haloes by studying intracluster supernovae. 

Because stellar haloes are dim and diffuse, a great deal of observational time is required to adequately measure their properties and characteristics. Therefore, simulations offer a useful alternative method of study, by providing predictions, within a rigorously constrained theoretical framework, about stellar halo formation that can be supported or refuted by observations, as well as probing aspects of the stellar halo that are not (yet) directly observable.

\citet{2019MNRAS.485.2589M} and \citet{2019MNRAS.484.4471F} have recently studied the stellar halo using the \textsc{Auriga} magnetohydrodynamic simulation. \citet{2019MNRAS.485.2589M} find that galaxies with a lower number of significant progenitors tend to have higher stellar halo masses and large negative metallicity gradients compared to those with a higher number of significant progenitors. \citet{2019MNRAS.484.4471F} find that accretions into the haloes of Milky Way Analogue (MWA) galaxies can produce highly eccentric, metal-rich stars in the stellar halo, similar to those observed by \citet{2019MNRAS.486..378L}.

These detailed hydrodynamical studies provide a great deal of information on the formation histories of stellar haloes, however there is also much to be gained from semi-analytic models. These models are better-suited to the study of large samples, and enable an analysis of the effects of different formation and chemical prescriptions in a wide range of halo masses to be run with minimal computational time. Results from semi-analytic models can also be compared to and informed by such highly detailed hydrodynamical simulations.

Therefore, in this work, we study the properties of stellar haloes surrounding galaxies and galaxy clusters in the \lgaltt{} semi-analytic model of galaxy evolution. Furthermore, we implement different prescriptions for the stellar stripping of satellite galaxies, and investigate their impact on the formation and chemical enrichment histories of stellar haloes.

This paper is organised as follows: In Section \ref{sec:Simulation}, we introduce \lgaltt{}, and describe the modifications made to the code. This section also describes the analysis methods used. Section \ref{sec:Results} provides our results and a comparison to those from other recent observations and simulations. In Section \ref{sec:Conclusions} we highlight and summarise our main conclusions.

%% file: Chapters/Simulation.tex

The \lgal{} semi-analytic model \citep{Springel+01,Springel+05,DeLucia&Blaizot07,Guo+11,Henriques+15,Henriques+20} runs on N-body simulations of dark matter (DM) structure formation, and self-consistently models the evolution of baryonic components in galaxies and DM subhaloes. The semi-analytic formalism works by modelling baryons as homogeneous spatial components, rather than as computational particles. This makes semi-analytic models such as \lgal{} highly efficient, running many thousands of times faster than hydrodynamical simulations of comparable scale.

Each model galaxy in \lgal{} consists of a central supermassive black hole (SMBH), stellar bulge, stellar disc and gas disc (each split into 12 concentric annuli, see below), circumgalactic medium (CGM), ejecta reservoir, and stellar halo. Mass and energy is then transferred between these components via differential equations describing key astrophysical processes such as gas cooling, star formation, stellar/SMBH feedback, and others.

As in hydrodynamical simulations, the free parameters in the model, which describe for example the efficiency of star formation or SMBH accretion, are tuned by fitting to a set of observational constraints. In \lgal{}, an MCMC formalism is used to simultaneously constrain model parameters to the observed galaxy stellar mass functions at $z=0-2$, quenched galaxy fractions as a function of stellar mass at $z=0-2$, and \textsc{Hi} mass function at $z=0$ \citep{Henriques+20}.

In this work, we choose to run \lgal{} on the \mil{} simulation \citep{Springel+05}, so that we can sample from a large number of well-resolved MWA and galaxy cluster systems. A Planck cosmology with a Hubble parameter of H$_0$ = 67.3 km s$^{-1}$Mpc$^{-1}$ is used, leading to a box side length of 714 Mpc (\ie{}480 h$^{-1}$ Mpc). The \mil{} data consists of 64 output snapshots. \lgal{} further simulates 20 timesteps between each snapshot. When using a representative sample of haloes from 22 tree files, the total number of galaxies in our sample was 1,151,404, without any selection criteria. Each of the 512 tree files in \lgal{} contains `merger trees' for the DM subhaloes in the \mil{} simulation identified using \textsc{Subfind} \citep{Springel+01}, upon which \lgal{} simulates the baryonic components.

Galaxies are defined as either `central' or `satellite' galaxies in \lgal{}. A central galaxy is the most massive galaxy in a `friends-of-friends group'. A friends-of-friends (FoF) algorithm finds groups of DM haloes according to a particular linking length. Any two DM haloes which are within this length are defined to be part of the same group \citep{Springel+05}. In the \mil{} simulation, this linking length is 20 per cent of the mean distance between simulation particles across the whole simulation box \citep{Davis+85}. This mean linking length is approximately 231.5$h^{-1}$ kpc \citep{Springel+05}.

The latest version of the model, \lgaltt{} \citep{Henriques+20}, is used for this study. This version includes the galactic chemical enrichment (GCE) prescription introduced by \citet{2013MNRAS.435.3500Y}, which tracks the lifetimes of stars and measures the mass and energy they deposit into the interstellar medium (ISM) and circumgalactic medium (CGM) over time. Eleven elements are tracked in this model, produced by thermonuclear (Type Ia) and core-collapse (Type II) supernovae (SNe), and winds from asymptotic giant branch (AGB) stars. These are H, He, C, N, O, Ne, Mg, Si, S, Ca, and Fe. Yields for AGB stars are taken from \citet{Marigo01}, those for SNe-Ia are from \citet{Thielemann+03}, and those for SNe-II are from \citet{Portinari+98}. The lifetimes of SN-Ia progenitors are governed by an analytic delay-time distribution (DTD), as described in \citet{2013MNRAS.435.3500Y} and in Section {\ref{sec:DTDs}} below.

\lgaltt{} also includes molecular hydrogen (H$_{2}$) formation \citep{Fu+10} and, following \citet{Fu+13}, the cold gas and stars in galaxy discs are divided into concentric annuli. This allows for a more physical approach when describing the radial distribution of material in galaxies, and is therefore more conducive to implementing the gradual stripping model for stars and cold gas discussed in Section \ref{sec:TidalDisruptions}.

An active galactic nucleus (AGN) feedback prescription is also implemented into \mbox{\lgaltt{}}, which reduces the cooling rate of gas onto galaxies in any system with a large enough supermassive black hole (SMBH) and hot gas reservoir, including Milky Way-mass systems, potentially.

A detailed description of the modelling techniques used in \lgaltt{} is provided by \citet{Henriques+20} and in the supplementary material.\footnote{\url{https://lgalaxiespublicrelease.github.io/Hen20_doc.pdf}}

\subsection{MWA and galaxy cluster samples}\label{sec:samples}
For our MWA analysis (Section \ref{sec:GalaxyResults}), the following selection criteria were used to select galaxies at $z=0$ from \mbox{\lgaltt{}}:

\begin{itemize}
    \item $2\times10^{10} \leq{} M_*/{\rm{\Msun}} < 8\times10^{10}$,
    
    \item 1.0 $\leq{}$ log(SFR/$\Msun$/yr) $<$ 5.0,
    
    \item ${M}\sub{*,disc}/({M}\sub{*,disc} + {M}\sub{*,bulge}) \geq{} 0.7$,
    
    \item Central galaxies only,
\end{itemize}
where SFR is the star formation rate of the galaxy. This returned 417 MWA systems, with stellar halo masses ranging from $2.09\times10^3\Msun$ to $3.26\times10^{10}\Msun$ and a mean of $4.02\times10^{9}\Msun$.

When comparing specifically to the MWA analysis in \textsc{Auriga} by \citet{2019MNRAS.485.2589M} in Section {\ref{sec:SigProgs}}, we adopt their simpler selection criterion of $1\times10^{12}<{M}_{200}/\Msun<2\times10^{12}$, where $M_{200}$ is defined as the mass within the radius $R_{200}$ enclosing a density 200 times greater than the critical density of the Universe. This returned 6 086 systems.

For our galaxy cluster analysis (Section \ref{sec:GalaxyClusters}), we select systems with a mass of $M\sub{200} > 10^{13}\Msun$. This selection returned 2 841 systems. The most massive galaxy cluster had a total mass of $M\sub{500}\approx2\times10^{15}\Msun$, where $M\sub{500}$ refers to the enclosed mass inside a radius of $R_{500}$.

\subsection{Stellar and cold gas stripping}\label{sec:TidalDisruptions}

Only satellite galaxies (\ie{}galaxies which live within the DM halo of a more massive galaxy, see Section {\ref{sec:Simulation}}) undergo stellar and cold gas stripping in \lgal{}. This stripping is defined as the transfer of stars and cold gas (\ie{}ISM) from the satellite galaxy to the stellar halo and CGM surrounding the central galaxy, respectively. This process is different to galaxy mergers, in which the satellite adds all its material to the central galaxy itself.

By default, \lgaltt{} uses an \textit{instantaneous} stellar and cold gas tidal stripping prescription (\ie{}total tidal disruption), hereafter the ``instantaneous stripping model'', which transfers all material from a satellite galaxy to the central's stellar halo and CGM as soon as the total disruption criterion is met: \ie{}when the satellite's DM subhalo falls below the \mil{} resolution limit, and its baryonic density is less than the average background DM density within the pericentre of its current orbit (see \citealt{Henriques+20}).

Alternatively, a \textit{gradual} stellar and cold gas tidal stripping prescription was developed by \citet{2010MNRAS.403..768H}, to better capture the tidal forces acting on satellite galaxies over time. In this work, we have incorporated a re-developed version of this ``gradual stripping model'' into \lgaltt{}, which calculates a stellar and cold gas tidal stripping radius ($r\sub{strip}$) centered on the satellite galaxy, beyond which all material is removed. This radius is given by

\begin{equation}\label{rdisrupt}
    r\sub{strip}=\frac{1}{\sqrt{2}}\frac{\sigma_{\rm{sat}}}{\sigma_{\rm{halo}}}r_{\rm{sat}},
\end{equation}

\noindent where $\sigma_{\rm{sat}}$ and $\sigma_{\rm{halo}}$ are the velocity dispersions of the satellite and central galaxy, respectively, and $r_{\rm{sat}}$ is the distance between the satellite galaxy and the centre of its host FoF group. This equation describes the radius at which the radial forces acting on the orbiting satellite (namely, the gravitational binding force, the tidal force, and the centripetal force) are balanced, under the simplifying assumption that both the central subhalo and the satellite subhalo have an isothermal mass profile.

In \citet{2010MNRAS.403..768H}, an exponential profile for the stellar and cold gas discs was assumed, and a $r^{1/4}$ profile for stellar bulges. The mass outside $r\sub{strip}$ was then calculated (assuming a flat metallicity gradient) and removed from the satellite accordingly.

In \lgaltt{}, galactic discs are instead divided into 12 concentric annuli (\ie{}`rings'), extending out to 40.5 kpc$h^{-1}$. Galaxy evolution processes, such as star formation and SN feedback, are then modelled self-consistently within each ring. This allows for a deviation from an assumed exponential density profile and flat metallicity gradient in discs. Therefore, in this work, these rings are used to refine the gradual stripping model. For satellites, only material in rings with radii larger than $r\sub{strip}$ is transferred to the stellar halo (for stars) and CGM (for cold gas). Should the stripping radius lie between two rings, the appropriate fraction of the ring is stripped assuming an even distribution of mass and metals within the individual ring. The average $r\sub{strip}$ value in \lgaltt{} for satellites surrounding Milky Way-mass galaxies at redshift 0 was found to be 4.1 kpc.

\begin{figure*}
	\includegraphics[width=\textwidth]{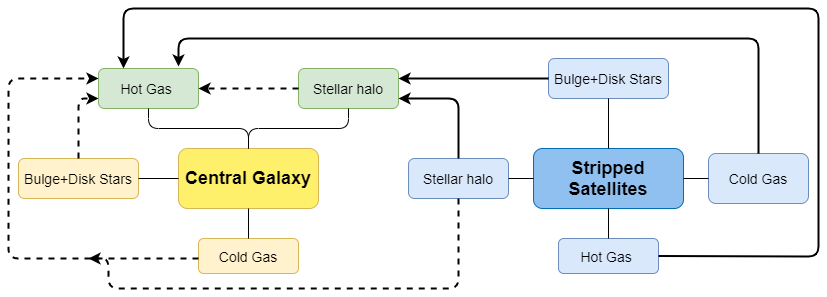}
    \caption{Solid arrows represent mass transfer when satellite stripping takes place. Dashed arrows represent hot gas enrichment sources. Hot gas enrichment from cold gas is due to reheated gas in outflows. Following stellar stripping, stellar halo stars in satellite galaxies which are accreted enrich the hot gas component of the central galaxy. Enrichment mechanisms in the satellite galaxies are not shown here for readability, but are present in the model.}
    \label{fig:SimFlow}
\end{figure*}

We note that the stripping models for stars and cold gas discussed here are distinct from the hot gas stripping model that is already incorporated into \lgaltt{}. Hot gas stripping, caused by tidal and ram pressure forces acting on the weakly bound CGM of satellite galaxies, is always calculated gradually in \lgaltt{}, as discussed in the Supplementary Material. The prescription described here is also distinct from the ram-pressure stripping model developed by \citet{Luo+16} to describe the removal of cold gas from satellites by hydrodynamical, rather than gravitational, effects.

We also note that, in \lgaltt{}, stellar haloes only form through this ex-situ process of satellite stripping. However, the possible impact of including in-situ halo formation processes into \lgaltt{} is discussed in Section \ref{sec:insitu}.

Fig. \ref{fig:SimFlow} provides a visual representation of the mass transfer involved in stripping events, as well as hot gas enrichment sources. In both satellite and central galaxies, the bulge, disc, and stellar halo stars enrich the hot gas component through outflows. The cold gas in the disc undergoes reheating and enriches the hot gas as well. Once the satellite galaxy undergoes stellar stripping, the stellar halo component is transferred to that of the central galaxy. Depending on the strength of the stellar stripping, both the disc and bulge can have stars stripped into the stellar halo of the central galaxy. These newly added stars would then begin to enrich the hot gas of this galaxy.

\subsection{Galaxy stellar mass function}\label{sec:SMF}

\begin{figure}
	\includegraphics[width=\columnwidth]{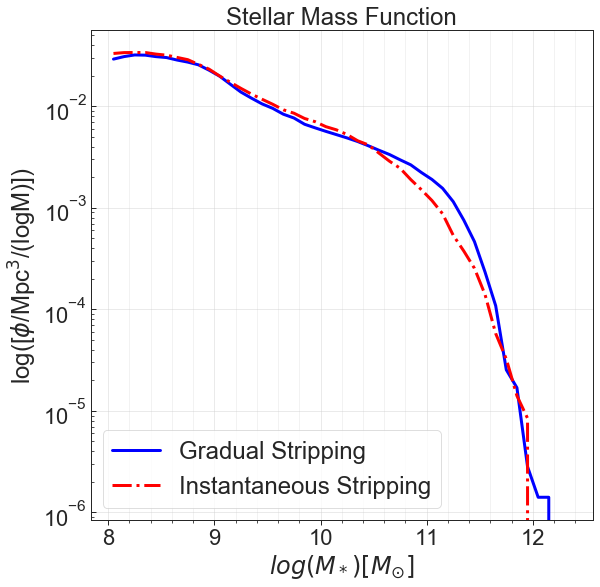}
    \caption{The stellar mass functions at $z=0$ produced by \lgaltt{} when the default instantaneous stripping model is used (red dot-dashed line), and when the gradual stripping model is used (solid blue line). The greater stripping efficiency of gradual stripping results in an increase in the stellar mass of L* galaxies.}
    \label{fig:SMF}
\end{figure}

Fig. \ref{fig:SMF} shows the galaxy stellar mass function (SMF) from \lgaltt{} at $z=0$ for the instantaneous stripping model (red dot-dashed line) and gradual stripping model (blue solid line). Gradual stripping results in an increase in L$^{*}$ galaxies (those at the knee of the SMF). This is likely due to the increased efficiency of the gradual stripping model in comparison to instantaneous stripping. We find that, for gradual stripping, more cold gas is removed overall from the satellite galaxy population. Therefore, when the satellite later merges with the central galaxy, less cold gas is available to feed the central galaxy's SMBH via quasar-mode accretion (see the \lgaltt{} supplementary material). This reduces the AGN feedback in the central galaxy, which, in turn, results in increased cooling and star formation rates in comparison to the instantaneous stripping model. As a result, by $z=0$, galaxies that were just below the knee of the SMF now have a greater stellar mass. 

This does not seem to affect galaxies with stellar masses above $\sim{}10^{11.5}\Msun$, however, possibly due to the high SMBH and hot gas masses in these systems making AGN feedback very efficient, regardless of the type of stripping model. 

\subsection{Parameter changes}\label{sec:parameterchanges}

When analysing the chemical evolution of galaxies in \lgaltt{}, \citet{Yates+21a} found that the default version of the model, which allows 70 per cent of the material ejected by SNe to directly enrich the ambient ISM, returned stellar and ISM metallicities that were too high compared to the latest global and spatially-resolved observational data (\eg{}\citealt{Zahid+17,Belfiore+17,Yates+20}). Consequently, the amount of direct ISM enrichment by SNe was decreased to $10-20$ per cent, with the remainder of the material ejected by SNe allowed to directly enrich the CGM surrounding galaxies. These values were motivated by observations and simulations showing that a significant amount of metals can be efficiently removed from galaxies via SNe-driven galactic outflows (\eg{}\citealt{Martin+02,Tumlinson+11,Li+17,Gatto+17}).

\citep{Yates+21a,Yates+21b} have shown that this adjustment brought much better agreement between this modified \lgaltt{} model and (a) the observed mass -- metallicity relations for stars and gas at $z=0$, (b) the temperature -- metallicity relation for nearby galaxy clusters, (c) the radial metallicity profiles in nearby disc galaxies, and (d) the metallicities observed in damped Lyman-$\alpha$ (DLA) systems back to $z\sim{}4.5$.

In this work, we also use this modified version of \lgaltt{}, known hereafter as \lgaltt{}-MM. In detail, the fraction of material released by SNe-II that can directly enrich the hot CGM ($f_{\rm{SNII,hot}}$) is increased from 0.3 to 0.9, the equivalent SNe-Ia fraction ($f_{\rm{SNIa,hot}}$) is increased from 0.3 to 0.8, and the equivalent AGB wind fraction ($f_{\rm{AGB,hot}}$) is increased from 0.0 to 0.25. Ultimately, these changes affect the metallicity of stars and gas in galaxies, as well as stellar halo metallicities following stellar stripping.

By increasing these parameters, a smaller fraction of metals produced by SNe-II, SNe-Ia, and AGB stars are deposited directly into the ISM. Consequently, the stars which form in satellite (and central) galaxies will have a lower metallicity, which in turn will lower the average metallicity of the stellar haloes which form from this material when it is stripped.

In addition, following \citet{Yates+21a}, the gas inflow velocity ($\alpha\sub{inflow}$) is decreased from 1000 to 600 km/s/Mpc and the fraction of stellar objects in the mass range $3-16 \Msun$ which produce SNe-Ia ($A\sub{SNIa}$) is decreased from 0.04 to 0.035. These two parameters are less relevant to this paper, but their reduction from their default values improves agreement with observations of [$\alpha$/Fe] in the stellar populations of early-type galaxies.

Initial work on the stellar halo metallicities of galaxies motivated the use of the \lgaltt{}-MM version of the model. In the default model, stellar halo metallicities in Milky Way-mass galaxies were found to be in worse agreement with the ensemble literature data compared to here \citep{geoff_grant_2020}.

Using the above GCE model, we track the chemical composition of stellar haloes, the mass in elements produced by SNe-II, SNe-Ia, and AGB stars in the stellar halo which are deposited into the hot gas component, and the mass in elements added to the hot gas by outflows from the central galaxy. 

\subsection{In-situ stellar halo formation}\label{sec:insitu}

Currently, \lgaltt{} does not incorporate a mechanism for in-situ stellar halo formation, either through the heating of thick disc stars or star formation in the halo itself. While many of the model results presented in this paper agree well with observations without the need for in-situ star formation, it is nevertheless an important process which should be considered when assessing our results in comparison to observational data.

In their study of \textit{Gaia} DR2 and APOGEE DR14 stars, \citet{2019A&A...632A...4D} find that heated thick disc stars (i.e. stars scattered from the central galaxy during mergers/accretion events) constitute 40 per cent of the stellar halo within 2-3 kpc of the Sun when selected as $\textrm{[Fe/H]} < -1$, with the remaining contribution being from accreted (\ie{}ex-situ) stars. When selecting halo stars in the solar neighbourhood based on kinematics, heated thick disc stars are expected to form the majority of the stellar halo.

When studying the metallicity distribution function (MDF) of stellar halo stars within $\sim{}5-10$ kpc from the Sun in SDSS Stripe 82 and building on the work of \citet{2007Natur.450.1020C,2010ApJ...712..692C}, \citet{2013ApJ...763...65A,2015ApJ...813L..28A} determine two chemically distinct components. \citet{2013ApJ...763...65A} find that the metal-richer (in-situ) component comprises $\gtrsim{}65$ per cent of the local stellar halo population, and \citet{2015ApJ...813L..28A} use a larger sample to find this component comprises 45-65 per cent.

Similar results are also found in some simulation studies. For example, in the \textsc{ARTEMIS} hydrodynamical simulation, \citet{Font+20} find that stellar halo and bulge stars born in-situ dominate the non-disc component of MWA galaxies within $\lesssim45$ kpc, with accreted stars dominating at larger radii.

Likewise, \citet{2019MNRAS.485.2589M} find that in-situ stars are the dominant stellar halo component within $\sim20$ kpc in many of their \textsc{Auriga} systems. Even at radii greater than 100 kpc, the in-situ component can form up to $10-20$ per cent of the total stellar halo mass. However, when compared to nearby galaxies, the authors find stellar halo masses larger than observed, and suggest that the in-situ component may be too strong in their model. Furthermore, in some model galaxies, in-situ stars result in drastic increases to the median iron abundance, particularly for radii $\lesssim30$ kpc (see their fig. 4).

In contrast, in the Aquarius hydrodynamical simulations used by \citet{Cooper+15}, in-situ stars constitute only 30-40 per cent of the total stellar halo mass. Of this in-situ component, heated stars constitute $\sim{}3-31$ per cent, with the rest formed from stripped or smoothly accreted gas in the halo.

Similarly, \citet{Pillepich+15} study in-situ and ex-situ star formation in a late-type spiral galaxy using the Eris hydrodynamical simulation. They find that the in-situ stellar component accounts for 25 per cent of the stellar mass of the inner halo ($R<20$ kpc), and only 3 per cent of the outer halo ($20<R<235$ kpc).
 
In galaxy clusters, in-situ stars can also contribute a small but non-negligible component to stellar haloes (\ie{}the ICS). For example, in their study of clusters in the TNG300 simulation, \citet{Pillepich+18b} find that in-situ stars contribute 30-40 per cent of the combined BCG+ICS mass for systems with $10^{13} < M\sub{200}/\Msun \lesssim 4\times10^{15}$ at $z=0$. At radii greater than 100 kpc, the ICS mass is consistently only around 10 per cent in-situ. Qualitatively similar results are also found for the \textsc{Hydrangea} simulation (Bah\'{e} et al., in prep.).

Because of the evident presence of an in-situ component in the above works, we make an assessment of the likely impact that including in-situ formation in \lgaltt{} may have on our results in Section {\ref{sec:Results}}.

%% file: Chapters/Analysis.tex
\subsection{Scaling of apertures}\label{sec:ProfileScaling}

For our galaxy cluster analysis (Section {\ref{sec:GalaxyClusters}}), it is important to scale model ICS measurements to the apertures used when observing real clusters, as L-Galaxies 2020 only provides total ICS masses within $R_{200}$. To do this, an assumed radial profile for the ICS mass density is required. In the case of conversions from $R_{200}$ to $R_{500}$, the Python package \textsc{Colossus} was used \citep{2018ApJS..239...35D}.

Different stellar halo density profiles have been discussed in the literature. For example, \citet{2018MNRAS.475..648P} assume a power-law in their study of groups and clusters, while \citet{2004ApJ...617..879L} and \citet{2015A&A...577A..19V} use a Navarro-Frenk-White (NFW, \citealt{Navarro+97}) profile. Both were tested in the \lgaltt{}-MM model.

The NFW profile is given by 

\begin{equation}\label{eq:NFW}
    \rho(r) = \frac{\rho_0}{(r/r\sub{s})(1+r/r\sub{s})^2} ,
\end{equation}

\noindent where $\rho_0$ is the normalisation and $r\sub{s}$ is the scale radius. The scale radius is found with $R_{200}$ and the concentration parameter, $c$,

\begin{equation}\label{eq:conc}
    r\sub{s} = \frac{R_{200}}{c} ,
\end{equation}

\noindent where $c$ itself is found using the fitting function derived by \citet{2004A&A...416..853D},

\begin{equation}\label{eq:concsolve}
    c(M,z) = \frac{c_0}{1+z}\left(\frac{M_{200}}{10^{14}h^{-1}\tn{M}_{\odot}}\right)^{\alpha} .
\end{equation}

\noindent In their $\Lambda$CDM model, \citet{2004A&A...416..853D} found a concentration normalisation of $c_{0} = 9.59$ and a slope of $\alpha = -0.102$. A Hubble parameter of $h=0.673$ is assumed in our work. 

This allows $r\sub{s}$ to be found, leaving only $\rho\sub{0}$ unknown. Integrating eq. \ref{eq:NFW} from the centre of the cluster to $R\sub{200}$ produces the total ICS mass, i.e. 

\begin{equation}\label{eq:mass}
    M\sub{ICS,tot} = 4\pi \int\sub{1pc}^{R_{200}} \frac{\rho_0 r^2}{(cr/r_{200})(1+cr/r_{200})^2} \tn{d}r .
\end{equation}

\noindent Rearranging allows for the density normalisation to be found for each galaxy cluster:

\begin{equation}\label{eq:norm}
    \rho_0 = \frac{M\sub{ICS,tot}}{4\pi} \left[\int\sub{1pc}^{R_{200}} \frac{ r^2}{(cr/r_{200})(1+cr/r_{200})^2} \tn{d}r\right]^{-1} .
\end{equation}

\noindent This allows the ICS mass inside any aperture to be found by replacing the upper integration limit in eq. \ref{eq:mass}. A lower limit of 1 pc is used here, though there is very little difference in the results when larger inner radii are used, given that the outer limits are very large.

%% file: Chapters/Results/GalaxyResults.tex
\subsubsection{Stellar halo iron abundances}\label{sec:FeH}

\begin{figure}
	\includegraphics[width=\columnwidth]{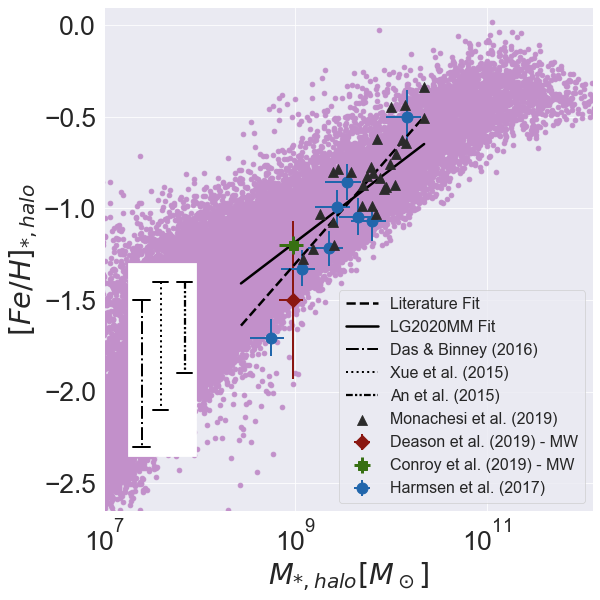}
    \caption{Mean stellar halo iron abundances within $R_{200}$ in \lgaltt{}-MM as a function of stellar halo mass (purple circles), when applying the gradual stripping model for stars and cold gas. This is compared to the observations of \citet{2019MNRAS.490.3426D} (dark red diamond), \citet{2019ApJ...887..237C} (green square), and \citet{2017MNRAS.466.1491H} (blue circle, measurements of nearby spiral galaxies), along with the \textsc{Auriga} results of \citet{2019MNRAS.485.2589M} (black triangles).} 
    \label{fig:FeHHT09}
\end{figure}

The stellar halo mass -- metallicity relation (\MZhR{}) at $z=0$ for \lgaltt{}-MM, when applying the gradual stripping model, is shown in Fig. \ref{fig:FeHHT09} (purple points). Stellar halo iron abundances, [Fe/H], in \lgaltt{} are calculated as follows,

\begin{equation}\label{eq:FeAb}
    \tn{[Fe/H]} = \tn{log}_{10}\left(\frac{M_{\rm{Fe}}}{M_{\rm{H}}}\right) - \tn{log}_{10}\left(\frac{M_{\rm{Fe},\odot}}{M_{\rm{H},\odot}}\right) ,
\end{equation}{}

\noindent assuming a solar photospheric iron abundance of $\tn{log}_{10}(\epsilon_{\rm{Fe},\odot})=7.50$ \citep{2009ARA&A..47..481A}. Both $M\sub{Fe}$ and $M\sub{H}$, along with other element masses, are tracked in \lgaltt{} for each baryonic component, including the stellar halo.

\begin{figure}
	\includegraphics[width=\columnwidth]{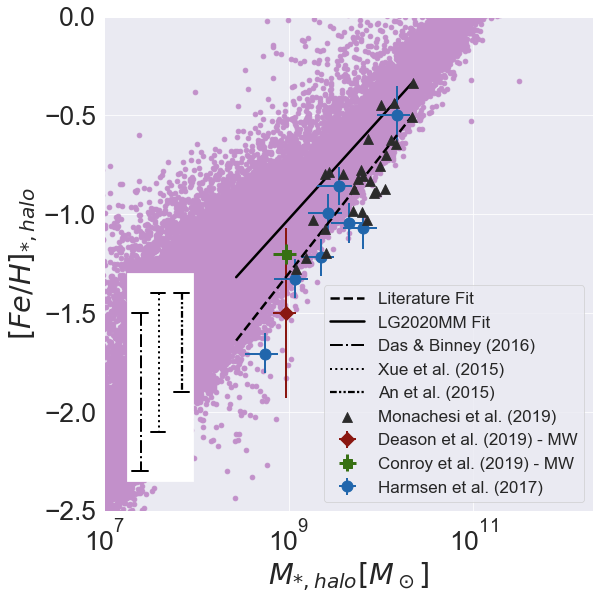}
    \caption{The same results as Fig. \ref{fig:FeHHT09}, but with the \lgaltt{}-MM metallicities re-scaled to be at 30\,kpc, using a simplistic linear profile. While this does not improve agreement, it does demonstrate that perhaps a more realistic profile could produce better results. The fitted linear slope is 0.60 for the combined literature sample, and 0.51 for \lgaltt{}-MM.}
    \label{fig:FeHHT09Scaled}
\end{figure}

\begin{figure}
	\includegraphics[width=\columnwidth]{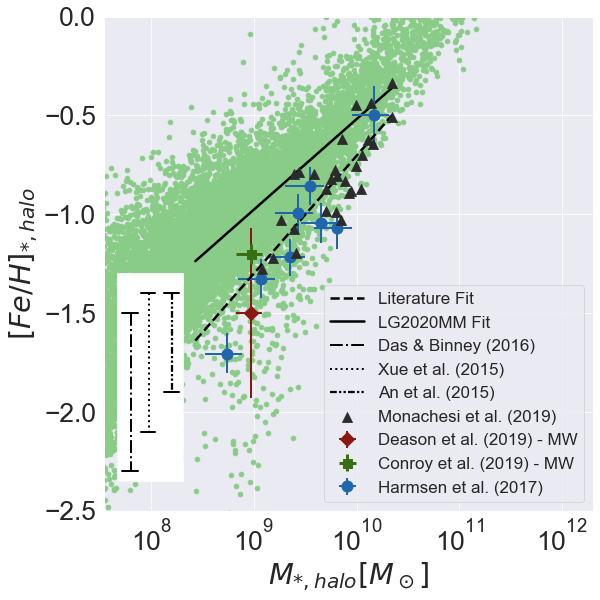}
    \caption{The \MZhR{} results when instantaneous stripping is used and the metallicities are re-scaled. As before, the literature fit produced a slope of 0.60, but the fit for the \lgaltt{}-MM results was 0.46, in poorer agreement with observations than the gradual stripping model.}
    \label{fig:FeHHT09Scaled_InstDisr}
\end{figure}

This is compared to one simulation data set and six observational literature data sets in Fig.~\ref{fig:FeHHT09}. \citeauthor{2019MNRAS.485.2589M} (2019, black triangles) measured \FeH{} at 30\,kpc for 27 Milky Way-size galaxies in the \textsc{Auriga} simulation. \citeauthor{2019MNRAS.490.3426D} (2019, brown diamond) infer the Milky Way's total stellar halo mass by observing red giant branch stars and adopt \FeH{} $\approx -1.5$, in qualitative agreement with the single-Gaussian fits to the observed stellar halo MDF from \citet{2013ApJ...763...65A} and \citet{2017ApJ...841...59Z}. \citet{2019ApJ...887..237C} use the \textsc{H3} survey to study the metallicity properties of the Milky Way's stellar halo, finding a mean metallicity of $\langle$[Fe/H]$\rangle \approx -1.2$. \citet{2017MNRAS.466.1491H} use red giant branch stars in the \textsc{GHOSTS} survey of nearby spiral galaxies to infer \FeH{} at a galactocentric radius of 30\,kpc. \citet{2016MNRAS.460.1725D} observe K giants in the Milky Way's stellar halo, inferring metallicity distributions for a metal-richer and a metal-poorer component. No assumptions of the Milky Way's stellar halo mass are made, so only their range of metallicities is provided in Figs. \ref{fig:FeHHT09}-\ref{fig:FeHHT09Scaled_InstDisr}. The upper limit of this range is the approximate peak metallicity of the metal-richer component, and the lower limit is the approximate peak of the metal-poorer component. Similarly, \citet{2015ApJ...809..144X} provide a range of metallicities by measuring K giants in the stellar halo at galactocentric radii between 10 and 80\,kpc. The upper limit of their range corresponds to the \textit{mean} of the metal-richer component, and the lower to the mean of the metal-poorer component. Finally, the peak metallicities of the metal-richer and metal-poorer components identified by \citet{2015ApJ...813L..28A} are also shown.

As is evident from much of this observational data, the MW's stellar halo is often considered to have two components, a metal-richer inner halo and a metal-poorer outer halo (see also Section {\ref{sec:Intro}}). The mean metallicity of the metal-poorer component was measured to be {\FeH{}} $= -1.9$, $-2.1$, and $-2.3$ by \citet{2015ApJ...813L..28A}, \citet{2015ApJ...809..144X}, and \citet{2016MNRAS.460.1725D}, respectively. The mean metallicity of the metal-richer component from these works was \FeH{} $= -1.4$, $-1.4$, and $-1.5$, respectively. These measurements are also in good agreement with the mean values obtained by \citet{2013ApJ...763...65A} and \citet{2017ApJ...841...59Z}.

There is very good agreement between the \lgaltt{}-MM \MZhR{} and the literature sample in terms of normalisation, however, there is a noticeable difference in slope. One possible reason for this is that the total stellar halo iron abundances are plotted for \lgaltt{}-MM in Fig.~{\ref{fig:FeHHT09}}, which include stars out to the virial radius of the halo, whereas multiple literature results measure the abundances only at 30\,kpc (\ie{}the inner halo). Therefore, a simple re-scaling of the stellar halo iron abundances in \lgaltt{}-MM to $r = 30$\,kpc is done, in order to determine if this could improve agreement.

Following the parameterisation derived by \citet{2019MNRAS.485.2589M} for \textsc{Auriga}, we assume a linear, negative radial profile for the stellar halo metallicity. While simplistic, this serves our purpose here, which is to check if aperture scaling is a possible method with which to increase the slope, thereby better matching literature data. Specifically, the form of the metallicity profile is $Z(r)=mr+C$, where $Z(r)$ is the metallicity as a function of galactocentric radius.

The mean slope of the metallicity profile determined by \citet{2019MNRAS.485.2589M} is $m=(-2.4\pm0.54)\times10^{-3}$dex\,kpc$^{-1}$ (and a median slope of $-2.95$). To find the constant $C$ for our \lgaltt{}-MM sample of MWAs, $Z(r)$ is integrated from 0 to $R_{\rm{vir}}$. This integral is assumed to equal the total stellar halo metallicity, \ie{}the value provided by \lgaltt{}-MM, allowing $C$ to be found for each system. Utilising this slope and galaxy-dependent normalisation, \Zhalo was calculated for model galaxies at 30\,kpc, and is shown in Fig. \ref{fig:FeHHT09Scaled}.

By applying such a metallicity profile, the overall slope of the model \MZhR{} relation increases, most noticeably for higher-mass systems. This is because the measurement of the metallicity is made at 30\,kpc in all cases, which, for massive haloes, is much closer to the metal-rich centre than the radius corresponding to the average value of \Zhalo{} plotted in Fig. \ref{fig:FeHHT09}.

A linear slope is a simplistic approach and the spread in slopes among the individual \textsc{Auriga} systems upon which our linear profile is calibrated is found to be large \citep{2019MNRAS.485.2589M}. Nevertheless, this analysis demonstrates the importance of considering stellar halo metallicity profiles when comparing semi-analytic models to observational data.

Fitting the combined literature \MZhR{} for stellar haloes returns $\tn{\FeH{}} = 0.60\,\tn{log}\sub{10}(M_{*\tn{,halo}}) - 6.66$. With no re-scaling of the metallicity profile, the gradual stripping model returns $\tn{\FeH{}} = 0.40\,\tn{log}\sub{10}(M_{*\tn{,halo}}) - 4.76$. Once the metallicity profiles are re-scaled, the slope of this relation moves closer to that seen in our literature data set, becoming $\tn{\FeH{}} = 0.51\,\tn{log}\sub{10}(M_{*\tn{,halo}}) - 5.66$. 

The \MZhR{} for \lgaltt{}-MM when applying the instantaneous stripping model was also checked, and is shown in Fig. \ref{fig:FeHHT09Scaled_InstDisr} (when again measuring metallicity at 30 kpc). This returned a fit of $\tn{\FeH{}} = 0.46\,\tn{log}\sub{10}(M\sub{*,halo}) - 5.10$, resulting in poorer agreement with our literature data set than our preferred re-scaled gradual stripping model. 

Finally, the dependence of the \MZhR{} on the SNe-Ia delay time distribution (DTD) was also checked. The three DTDs considered in \lgaltt{} are discussed in detail in Section {\ref{sec:DTDs}}. Here, we note that agreement between the model \MZhR{} relation of Fig. {\ref{fig:FeHHT09Scaled}} and the literature data set is slightly improved by the use of a Gaussian DTD, but worsened with a bi-modal DTD. This is in comparison to \lgaltt{}'s default power-law DTD. Table {\ref{table:fits}} summarises the slopes and normalisations of the fit to literature data and the model fits when assuming these three different DTDs.

\begin{table}
\centering
\begin{tabular}{c|cc|}
  \hline
  & Slope & Normalisation \\
  \hline
  Literature & 0.60 & -6.66 \\
  Bimodal & 0.50 & -5.45 \\
  Power-Law & 0.51 & -5.66 \\
  Gaussian & 0.56 & -6.07 \\
  \hline
\end{tabular}
\caption{The slope and normalisation values of the fits to the \MZhR{} relation for the literature sample, as well as to the relation when the three DTD models are used, where the power-law DTD is the default.}
\label{table:fits}
\end{table}

As a side note, \citet{2017MNRAS.466.1491H} measure stellar halo masses between 10-40\,kpc from the galactic centre and multiply by three in order to infer the total stellar halo mass. This was repeated in the model by distributing the total stellar mass with an NFW profile, and following the authors' method. This made very little difference to the \lgaltt{}-MM results, suggesting that this observational method returns accurate total stellar halo masses given our model. 

However, \citep{2010ApJ...712..692C} infer radial density profiles for both the inner and outer stellar halo in their study of this component's structure and kinematics. Future work on the stellar haloes in \lgal{} could adopt these profiles and radially distribute stellar halo mass in a manner motivated by observations. This would lead to fairer comparisons with literature masses.

The inclusion of in-situ halo stars in \lgaltt{}-MM would likely increase total stellar halo masses by at least $\sim20$ per cent in MWAs (\eg{}\citealt{Cooper+15,Pillepich+15,2019MNRAS.485.2589M,Font+20}). Mass-weighted average iron abundances would also likely increase, although to a lesser extent than stellar halo masses. This is because in-situ components tend to be more metal-rich than ex-situ components at radii $<30$ kpc (\eg{}\citealt{2019MNRAS.485.2589M}), as these stars would either originate from the relatively metal-rich stellar disc of the central galaxy or from CGM material which in \lgaltt{}-MM is highly enriched at early times \citep{Yates+21b}. Overall, such increases in stellar halo mass and [Fe/H] could lead to an improvement in the agreement between the \lgaltt{}-MM \MZhR{} relation and that observed.

Another possible observational result to compare to is the MDF of the Milky Way stellar halo. The observable MDF has been studied in many recent works (\eg{}\citealt{2007Natur.450.1020C,2010ApJ...712..692C,2013ApJ...763...65A, 2015ApJ...813L..28A, 2014A&A...568A...7A, 2015ApJ...809..144X, 2016MNRAS.460.1725D, 2020MNRAS.492.4986Y,2021ApJ...908..191C}). However, these observations measure the metallicities of individual stars within a fixed heliocentric distance (between $\sim5$ kpc and a few tens of kpc). \lgaltt{}, on the other hand, provides the average metallicity of the whole stellar halo. Consequently, the observed MDF within a sub-region of the Milky Way cannot be readily compared to \lgaltt{}. Additionally, the colour-magnitude selection criteria used to create spectroscopic surveys mean that each star has a probability of selection that is a function of age, metallicity, and distance. A lack of accounting for such selection effects will bias comparisons between observations and simulations (see \eg{}\citealt{2016MNRAS.460.1725D}).

Given these theoretical and observational issues, we choose not to make a detailed study of the shape of stellar halo MDFs from \lgaltt{}-MM in this work.

In conclusion, we find that \lgaltt{}-MM agrees well with the observed, low-redshift mass -- metallicity relation for stellar haloes, when accounting for the limited aperture of most observations. This further extends the good agreement seen for this model, which has already been shown to match the mass -- metallicity relations for the ISM and stars in galaxies \citep{Yates+21a}.

\subsubsection{Hot gas chemistry}\label{Disruption Events}

\begin{figure*}
	\includegraphics[width=0.8\textwidth]{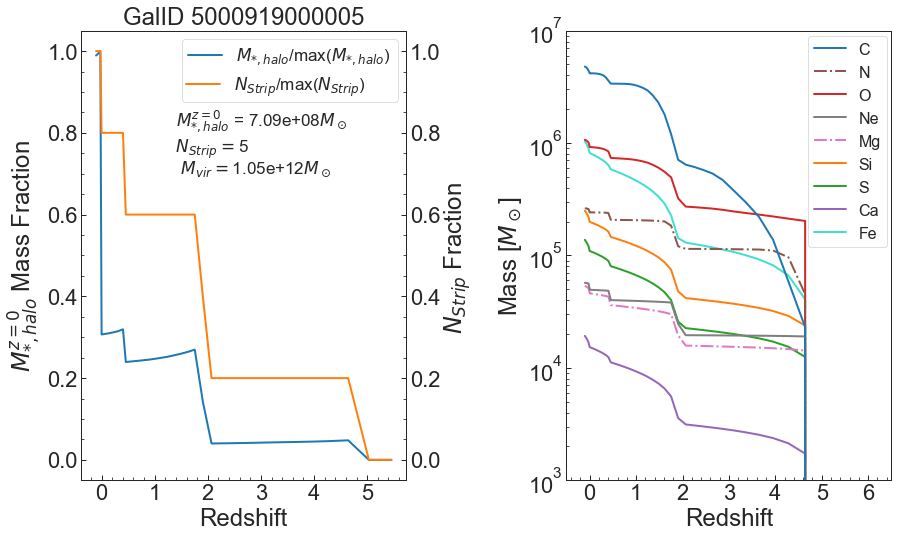}
    \caption{Left: The increase in stellar halo mass across the evolution of the MWA galaxy with galaxy ID 5000919000005 (blue line), along with the increase in the number of stellar stripping events (orange line), both plotted as a fraction of their maxima. Note that at $z\approx2$, two stellar stripping events occur, one at $z=2.07$, and one immediately after at $z=1.9$. Right: The cumulative mass in each element (excluding H and He) added to the hot gas by stellar halo SNe and AGB stars for this galaxy.}
    \label{fig:MWANDis}
\end{figure*}

\lgal{} provides an opportunity to study how stellar haloes enrich the surrounding hot gas over time. An example of the mass growth and enrichment evolution of a characteristic MWA stellar halo in \lgaltt{}-MM with the gradual stripping model applied is shown in Fig. \ref{fig:MWANDis}. The general trends seen for this example system hold across the entire MWA sample. One exception is that the stellar halo of this particular MWA formed at $z\sim{}4.6$, earlier than most other MWAs, which typically form at a mean redshift of $1.24$. This particular galaxy is chosen as it provides greater insight into the evolution of the enrichment from the stellar halo. 

The left-hand panel of Fig. \ref{fig:MWANDis} shows both the evolution of the stellar halo mass and number of stellar stripping events, $N\sub{strip}$, as a fraction of their maxima. The latter value simply tracks whether a stripping event occurred at a particular redshift, and whether any stellar mass was transferred to the stellar halo of the central galaxy from a satellite galaxy. The $z=0$ values are also quoted. We note that there are two stellar stripping events occurring in consecutive snapshots around $z\sim{}2$. The stellar halo mass shows a gradual decrease after each stripping event most noticeably at $z\sim{}0.5-1.9$. This is due to a combination of ejected material and stellar remnants being removed from the stellar halo component in \lgaltt{}-MM, as AGB stars shed their outer layers and SNe explode in the stellar halo.

The right-hand panel of Fig. \ref{fig:MWANDis} shows the cumulative element masses produced by stellar halo stars which are deposited into the surrounding hot gas. The rate of enrichment of each element changes with time, more drastically for some than others, for example carbon compared to nitrogen (see below). Specifically, we find that by $z=0$, carbon, oxygen, and iron were the most-ejected elements from stellar halo stars into the CGM for MWA systems.

\begin{figure*}
	\includegraphics[width=0.9\textwidth]{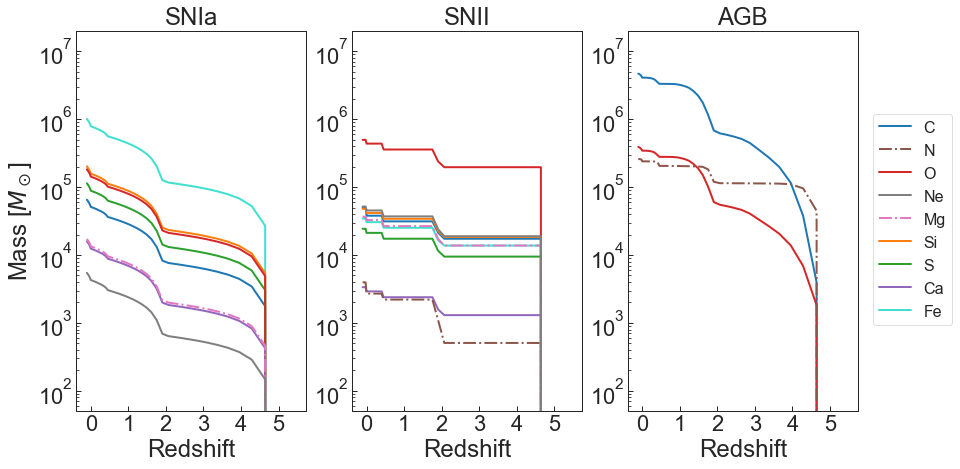}
    \caption{The cumulative stellar halo enrichment shown in the right panel of Fig. \ref{fig:MWANDis}, split into the individual enrichment channels: SNe-II, SNe-Ia, and AGB stars. Only newly-synthesised material is considered in this plot, excluding material already present at the formation of the SNe progenitors and AGB stars that doesn't get processed.}
    \label{fig:MWASplit}
\end{figure*}

\begin{figure*}
	\includegraphics[width=0.8\textwidth]{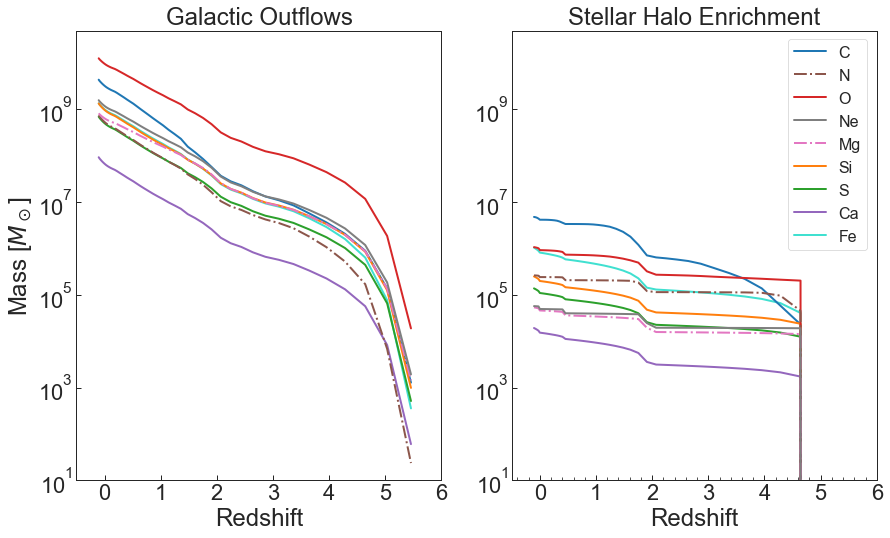}
    \caption{A comparison between the cumulative hot gas enrichment via galactic outflows from the central galaxy (left) and from stellar halo stars (right) for the MWA system shown in Fig. \ref{fig:MWANDis}.}
    \label{fig:MWAOutflows}
\end{figure*}

Fig. \ref{fig:MWASplit} again shows the chemical enrichment of the CGM by stellar halo stars as a function of redshift, but split into the three enrichment channels modelled in \lgaltt{} (\ie{}SNe-Ia, SNe-II, and AGB winds). By plotting these enrichment mechanisms separately, it was found that the high levels of oxygen present in the hot CGM gas at early times are mainly due to SNe-II whose progenitor stars were very recently accreted from stripped satellites (noting that there is no in-situ star formation in stellar haloes in \lgaltt{}). It is only at $z = 2$ that SNe-II progenitors are accreted again, with similar events occurring at later times. 

Once the majority of SNe-II have promptly exploded, carbon begins to be the dominant element ejected by halo stars, via AGB winds. AGB stars are lower in mass than SNe-II progenitors, hence it takes longer for them to evolve to the point where they shed large amounts of metal in their winds. An accretion event at $z = 2$ adds more AGB stars to the halo, increasing the rate of carbon production. By $z=0$, the enrichment from AGB stars was, on average, ten times stronger than from SNe-II, and three times stronger than SNe-Ia.

Out of the three populations, SNe-Ia produce the most iron. SNe-Ia progenitor lifetimes are determined by a DTD in \lgal{}. The default power-law DTD assumed in Fig. \ref{fig:MWASplit} allows 52 per cent of SNe-Ia to explode $>400$ Myr after star formation, meaning that a large amount of this iron is ejected significantly after any enrichment by SNe-II from the same stellar population. In Section {\ref{sec:DTDs}}, the impact that changing the SNe-Ia DTD has on chemical enrichment from halo stars is discussed, by comparing the default power-law DTD to a bi-modal DTD with a larger prompt component and a Gaussian DTD with a larger delayed component.

When focusing on AGB stars, we see that nitrogen is the most abundant element ejected at very early times ($z\gtrsim4.5$). In the AGB yield tables from \citet{Marigo01} used by \lgal{}, low metallicity AGB stars with mass $>4\Msun$ undergo hot-bottom burning and produce more nitrogen and oxygen than carbon, while those with mass $<4\Msun$ produce more carbon than nitrogen and oxygen. As such, the faster evolving high mass populations are responsible for the abundance of nitrogen at early times. The low mass, slower evolving populations produce the majority of the carbon at later times, leading to a higher rate of carbon enrichment compared to nitrogen enrichment in the periods between high-redshift stellar stripping events. Lastly, AGB stars with high metallicities and masses $>4\Msun$ produce similar masses of these three elements (see \citealt{2013MNRAS.435.3500Y}).

We note that all the yield sets used in the current versions of \lgaltt{} assume single star evolution models. Future work will look at the effect of binary star yields on the chemical composition of galaxies and stellar haloes in \lgal{}. This could be significant, particularly for SNe-II, since a large fraction of massive O stars ($\gtrsim{}70$ per cent) could evolve in binaries \citep{Sana+12}.

Lastly, Fig. \ref{fig:MWAOutflows} compares the cumulative hot gas enrichment from the stellar halo with that from the central galaxy via galactic outflows. Galactic outflows enrich the hot CGM gas much more significantly than the stellar halo in MWAs, adding total metal masses in the range $\sim10^8-10^{10}\Msun$ by $z=0$, while the stellar halo adds total metal masses in the range of $\sim10^4-10^7\Msun$. These values reflect the sum of the mass of all elements tracked in the model which are heavier than H and He.

Across all MWAs, the stellar halo contributes $0.1$ per cent of the total CGM enrichment at redshift 0, on average. A minimum value of $0.03$ per cent, and a maximum of $0.68$ per cent  was found. These small values are due to the central galaxy itself having ongoing star formation, and therefore an ongoing, high rate of enrichment of the CGM via outflows, especially in \lgaltt{}-MM which includes a high direct CGM enrichment efficiency for SNe. The contribution from MWA satellites is very minor. By $z=0$, satellites only contribute to $0.05$ per cent of the final CGM mass through outflows.

The enrichment from the galaxy also starts earlier, at $z\sim5.5$. Enrichment from the stellar halo can only begin after the first satellites are stripped, which occurs at $z\sim4.6$ for the particular system shown here. Furthermore, oxygen is always the dominant element in galactic outflows. This is, again, due to the ongoing star formation in the galaxy, which produces SNe-II progenitors continuously.

In conclusion, while the stellar halo provides an important component of hot gas enrichment, it is clearly sub-dominant in comparison to outflows for MWAs. However, for massive cluster galaxies ($M\sub{vir}\sim10^{14}\Msun$), we find that ICS enrichment can be comparable to, or even exceed, the mass produced by outflows from the central galaxy (see Section \ref{sec:GalaxyClusters}).

\begin{figure*}
	\includegraphics[width=.95\textwidth]{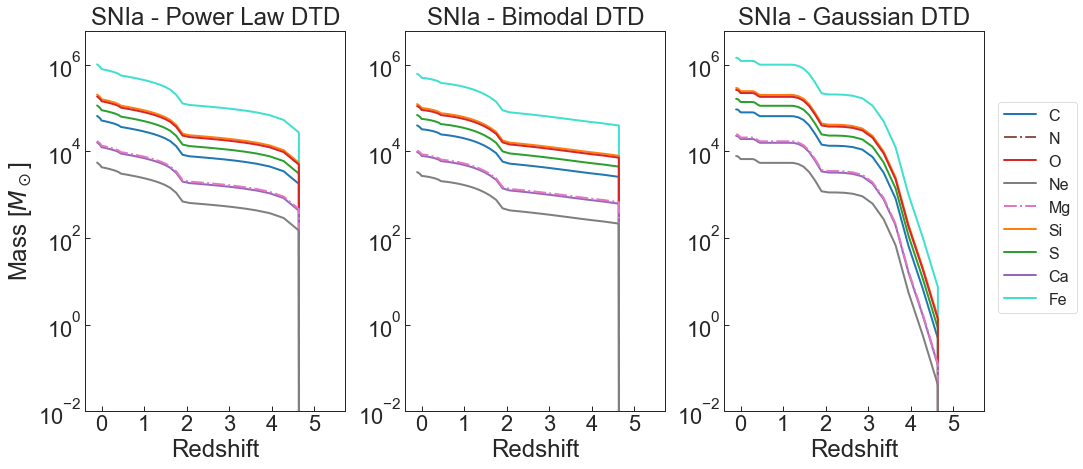}
    \caption{Enrichment characteristics from stellar halo SNe-Ia in three DTD models: The default power-law DTD, a bimodal DTD, and a gaussian DTD.}
    \label{fig:DTDSNIa}
\end{figure*}

\subsubsection{SNe-Ia Delay Time Distributions}\label{sec:DTDs}

The enrichment from SNe-Ia in stellar haloes when assuming different delay time distributions (DTDs) was also investigated. The DTD determines the rate of SNe-Ia events occurring after the birth of a given stellar population. Three different DTDs are currently implemented into \lgaltt{}. The default is the power-law DTD with a slope of $-1.12$ derived by \citet{Maoz+12}, which allows $\sim{}23$ per cent of SNe-Ia to explode within 100 Myr. The second is the bimodal distribution from \citet{2006MNRAS.370..773M}, wherein $\sim{}54$ percent of the SNe-Ia events occur within 100 Myr. Finally, a narrow Gaussian DTD centered on 1 Gyr (\ie{}with no prompt component) is considered, motivated by the study of \citet{Strolger+04}. These three DTDs nicely bracket a broad range of possible mean SNe-Ia delay times (see \citealt{2013MNRAS.435.3500Y}).

Fig. \ref{fig:DTDSNIa} shows the cumulative enrichment of the CGM by stellar halo SNe-Ia using these three DTDs. The power-law and bimodal distributions show similar results, except that there is slightly more enrichment from the bimodal DTD within the first $\sim{}100$ Myr after star formation. There are also mass differences by $z=0$ at approximately the same level for all elements.

The Gaussian DTD shows the most difference, as half of the SNe-Ia events occur only after 1 Gyr, well after the majority of SNe-Ia explode when assuming the power-law and bimodal DTDs. As such, the initial enrichment is much lower for the Gaussian DTD. By $z=2$, however, the total enrichment exceeds the other two models, and by $z=0$ the mass in all elements is highest when using the Gaussian DTD. This increased total enrichment is likely due to indirect changes to the galaxy evolution in \lgaltt{}-MM. The timing of enrichment from SNe-Ia, in combination with when stellar stripping events occur, could ultimately affect the final masses, \ie{}whether SNe-Ia enrichment occurs before or after stellar stripping in the three different DTD models. Furthermore, changing the DTD changes the distribution of SN feedback energy injection over time, to an extent.

\subsubsection{Significant progenitors}\label{sec:SigProgs}

\begin{figure*}
	\includegraphics[width=1.85\columnwidth]{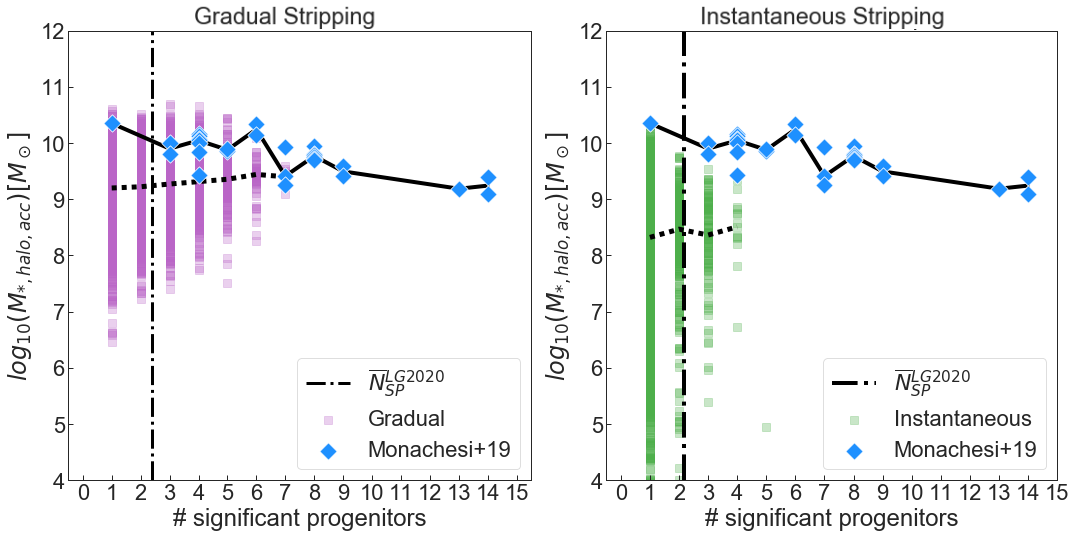}
    \caption{Left: The number of significant progenitors versus the accreted stellar halo mass for Milky Way-mass galaxies in the \lgaltt{}-MM model compared to the magneto-hydrodynamic results of \citet{2019MNRAS.485.2589M}, with the median stellar halo mass at each $N\sub{SP}$ given for both: a solid line for the literature data, and a dotted line for the \lgaltt{}-MM data. Black dot-dashed lines denote the mean $N\sub{SP}$ for all of our model galaxies. In \lgaltt{}-MM, the total stellar halo mass is used, as accretion is the only method of stellar halo formation in the model. \citet{2019MNRAS.485.2589M} provides stellar halo masses which include accretion and in-situ contributions, as well as accretion only. We use the latter, here. In gradual stellar stripping, a maximum of seven significant progenitors is found. $\overline{{N}}\sub{SP}$ is positioned at 2.40. Right: The same analysis using the default instantaneous stellar stripping method in \lgaltt{}-MM. There is a decrease in significant progenitor numbers as well as total stellar halo mass. A maximum of five significant progenitors is found, and $\overline{{N}}\sub{SP}$ is positioned at 2.18.}
    \label{fig:SigProgsHT09}
\end{figure*}

An interesting quantity to study when assessing the ex-situ formation of the stellar halo is the number of `significant progenitors' ($N\sub{SP}$). This is the number of stripped satellites which contribute 90 per cent of the final stellar halo mass, when ranked by the amount of mass they each contribute. \citet{2019MNRAS.485.2589M} study $N\sub{SP}$ in the \textsc{Auriga} simulation, which re-simulated galaxies (and their immediate surroundings) with virial masses $1\times10^{12}<{M}_{200}/\Msun<2\times10^{12}$. In order to make a similar study in \lgaltt{}-MM, we apply the same virial mass cut to our MWA sample in this section. Fig. \ref{fig:SigProgsHT09} shows the accreted stellar halo mass as a function of $N\sub{SP}$ for the gradual stripping model (left panel) and the instantaneous stripping model (right panel). The solid black line represents the median trend from \textsc{Auriga}, and the black dashed line represents the median trend from \lgaltt{}-MM.

Fig.~{\ref{fig:SigProgsHT09}} shows that the average $N\sub{SP}$ for galaxies in this virial mass range is very similar for the gradual stripping model and instantaneous stripping model, being $\overline{{N}}\sub{SP}=2.40$ and 2.18, respectively. However, the median accreted stellar halo mass is consistently higher in the gradual stripping model, \ie{}gradual stripping of stars produces more massive stellar haloes, on average. This is because, in the instantaneous stripping model, stripped stars are only transferred to the stellar halo once the satellite galaxy meets the total disruption criterion (see Section {\ref{sec:TidalDisruptions}}). However, in the gradual stripping scenario, satellites can be partially stripped before the total disruption criterion would have been satisfied, leading to a greater amount of accretion onto the stellar halo overall.


This enhanced stripping is also the cause of the larger maximum $N\sub{SP}$ seen for the gradual stripping model ($N\sub{SP,max}\sim{}7$, compared to $\sim{}4$ for the instantaneous stripping model). More satellite galaxies are able to contribute to the stellar halo in the gradual stripping model, although a smaller fraction of these contribute all of their stars. In other words, a low mass satellite which has almost all of its stars stripped can make a comparable mass contribution to a more massive satellite which only has its outer disc stripped.

When comparing to the \citet{2019MNRAS.485.2589M} study, both stripping models in \lgaltt{}-MM predict a lower $N\sub{SP,max}$ than the maximum of fourteen found in \textsc{Auriga}. Additionally, the \textsc{Auriga} results show a slight decrease in $M_{*\tn{,halo,acc}}$ with increasing $N\sub{SP}$, whereas the median trend in the \lgaltt{}-MM models is relatively flat.

Nonetheless, our results are in good agreement with results from Milky Way observations, which find that the majority of the inner stellar halo was formed from the accretion of one massive satellite galaxy, Gaia Enceladus, with a stellar mass of $\sim{}2.4 \times 10^9 \Msun$ \citep{2018Natur.563...85H}. 

A comparison with observations was also done for our MWA sample detailed in Section {\ref{sec:samples}} (\ie{}not  restricting the DM halo mass to $1\times{}10^{12} < M_{200}/\Msun{} < 2\times{}10^{12}$, but making selections using the stellar mass, SFR, and disc and bulge masses). At redshift 0, and across the MWA sample, there was a mean stellar halo mass of $4.02\times10^9\Msun$, with an average of 2.5 significant progenitors. 19.2 per cent of MWAs had one significant progenitor, contributing $2.33\times10^9\Msun$ to the stellar halo, on average.

The stellar halo masses for this MWA sample also agree with estimates of the Milky Way's stellar halo mass. For example, \citet{2019MNRAS.490.3426D} estimate $1.4\pm0.4\times10^9\Msun$, while \citet{2017MNRAS.466.1491H} assume a total MW stellar halo mass of $5.3\pm0.5\times10^8\Msun$. As stated above, our mean stellar halo mass for this sample is $4.02\times10^9\Msun$, but our results show a number of systems with a similar galactic stellar mass -- stellar halo mass relation. This is a notable contrast to magneto-hydrodynamic results, which tend to produce larger MWA stellar halos than observed, even when only considering the ex-situ component (as is done in \lgaltt{}-MM). For example, \citet{2019MNRAS.485.2589M} find a minimum total stellar halo mass (including ex-situ and in-situ stars) of approximately $7.4\times10^9\Msun$.

Finally, we note that the results presented in Fig.~{\ref{fig:SigProgsHT09}} would not change with the inclusion of an in-situ component, as the data shown from both \lgaltt{}-MM and \textsc{Auriga} only consider accreted stars. \citet{2019MNRAS.485.2589M} do, however, provide measurements which include in-situ stars.

%% file: Chapters/Results/GalaxyClusterResults.tex
\subsubsection{ICS masses}\label{sec:ICSMasses}

In the same manner that stellar haloes provide information on the interaction history of galaxies like the Milky Way, the ICS component can provide similar information for galaxy clusters. For example, \citet{2015MNRAS.448.1162D} find that negative ICS metallicity gradients could arise from tidal stripping of L$^*$ galaxies, given that these tend to have negative stellar disc metallicity gradients themselves, and the lowest metallicity stars on the outskirts are stripped first.

Observations also suggest that the ICS contributes significantly to its environment. For example, the ICS could produce as much as $\sim$25 per cent of the total metals found in the intracluster medium (\citealt{2009ApJ...691.1787S}, but see Section \ref{sec:ICSMassFractions}). 

\begin{figure}
	\includegraphics[width=\columnwidth]{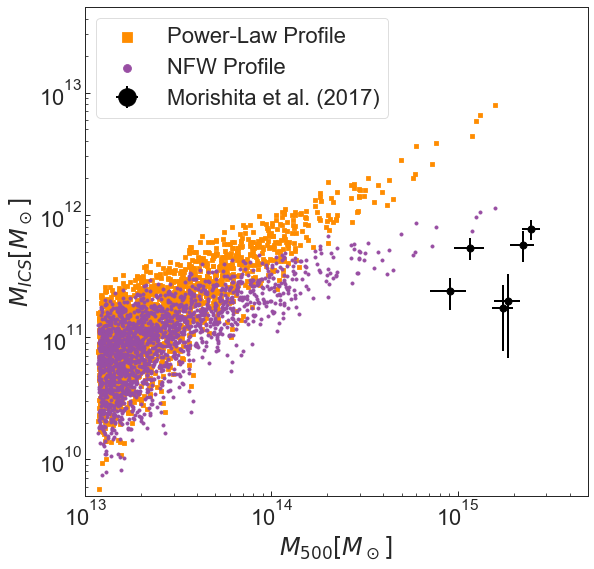}
    \caption{The ICS mass measured in galaxy clusters when using an aperture of 300\,kpc at a redshift of approximately $z=0.42$, which is the mean redshift of the \citet{2017ApJ...846..139M} data. Orange squares indicate \lgaltt{}-MM results (assuming the gradual stripping model) when the ICS is rescaled using a power-law profile, and the purple points are for the NFW profile.}
    \label{fig:MICSM500}
\end{figure}

Fig.~\ref{fig:MICSM500} shows the total ICS mass within 300\,kpc as a function of $M_{500}$ for \lgaltt{}-MM groups and clusters, when assuming the gradual stripping model. This relation is shown when assuming either a power-law (orange squares) or NFW (purple points) profile for the mass density distribution (see Section \ref{sec:ProfileScaling}). For comparison, six Hubble Frontier Field galaxy clusters at $z\sim{}0.42$ with $M_{500}\gtrsim{}10^{15}\Msun$ from \citet{2017ApJ...846..139M} are shown, where the ICS mass was measured by fitting the light profile within an aperture of 300\,kpc.

\citet{2017ApJ...846..139M} reconstruct the intracluster light of these six galaxy clusters using a new light profile fitting method, from which stellar mass profiles are created. The ICS was found to make up 5-20 per cent of the total cluster stellar mass, lower than is observed in the local Universe, suggesting that these ICS components at $z\sim{}0.42$ are still undergoing formation.

Fig.~\ref{fig:MICSM500} shows that, at fixed $M_{500}$, the \lgaltt{}-MM ICS masses are larger than observed by \citet{2017ApJ...846..139M}, when using the gradual stellar stripping model. A possible reason for this result is discussed later in the section. The observational data does show significant scatter in ICS mass, perhaps due to the difficulty associated with the observation of this component. Nonetheless, the NFW profile returns $M\sub{ICS}$ values for massive clusters which are closer to those observed by \citet{2017ApJ...846..139M}. The power-law profile distributes a large amount of mass towards the centre of the cluster, resulting in measured masses which can be nearly an order of magnitude larger within 300\,kpc. 

\begin{figure}
	\includegraphics[width=\columnwidth]{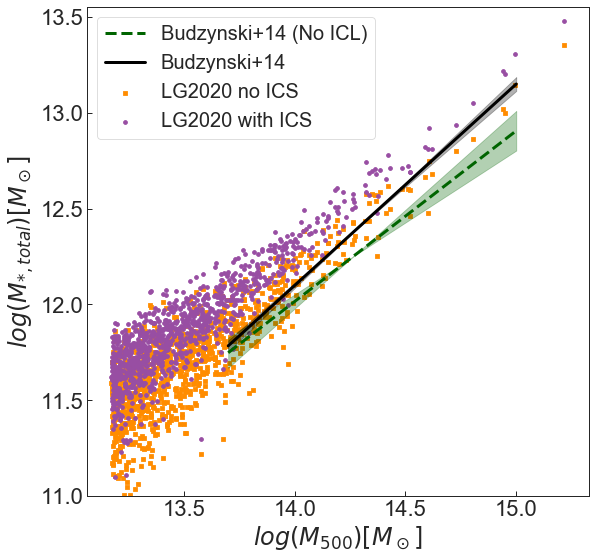}
    \caption{The total stellar mass of the cluster compared to \citet{2014MNRAS.437.1362B}, which provides data including and excluding measurements of the ICL using an aperture of $R_{500}$ at a mean redshift of 0.275. The purple points are the \lgaltt{}-MM data including the ICS mass, and the orange squares exclude the ICS mass.}
    \label{fig:MStellarHT09}
\end{figure}

In Fig. \ref{fig:MStellarHT09}, we show the total stellar mass contained in groups and clusters from \citet{2014MNRAS.437.1362B}, obtained by fitting the light profiles of 20 171 SDSS systems with $M_{500}\gtrsim{}10^{13.7}\Msun$ using an aperture of $R_{500}$, in the redshift range 0.15 $\lesssim z \lesssim$ 0.4. The authors bin the results by $M_{500}$ to find the mean properties in each bin. Results including and excluding the ICL are provided. \lgaltt{}-MM masses including the ICS are shown as purple points, and those excluding the ICS are shown as orange points. In both cases, we assume an NFW profile, as is done by \citet{2014MNRAS.437.1362B} in their modelling (see their section 3.2).

Fig.~{\ref{fig:MStellarHT09}} shows that \lgaltt{}-MM also slightly over-predicts the total stellar masses in groups and clusters when assuming the gradual stripping model. This is even true when excluding the ICS component, suggesting that either the brightest cluster galaxy (BCG) or satellite masses are also over-estimated relative to the observational data considered here.

In contrast, total cluster masses are found to match the \citet{2014MNRAS.437.1362B} data well when assuming the instantaneous stripping model, highlighting the smaller overall accretion onto the stellar halo that occurs in that model. However, cluster masses excluding the ICS are similarly over-predicted.

The agreement between observations and \lgaltt{}-MM in Figs. {\ref{fig:MICSM500}} and {\ref{fig:MStellarHT09}} would likely not be improved by the inclusion of an in-situ component. An increase in ICS mass of $\sim{}10-20$ per cent (see \citealt{Pillepich+18b}, Bah\'{e} et al. in prep.) would further increase the normalisation of the model trends in Fig. {\ref{fig:MICSM500}} away from the \citet{2017ApJ...846..139M} data. The strength of tidal stripping of stars and cold gas in \lgaltt{}-MM would likely need to be reduced in order to counteract this. In Fig.~{\ref{fig:MStellarHT09}}, the inclusion of an in-situ component would lower the orange points slightly, bringing \lgaltt{}-MM into better agreement with the \citet{2014MNRAS.437.1362B} data for ``no ICS'', but the total cluster masses (purple points) would remain unchanged.

\begin{figure}
	\includegraphics[width=\columnwidth]{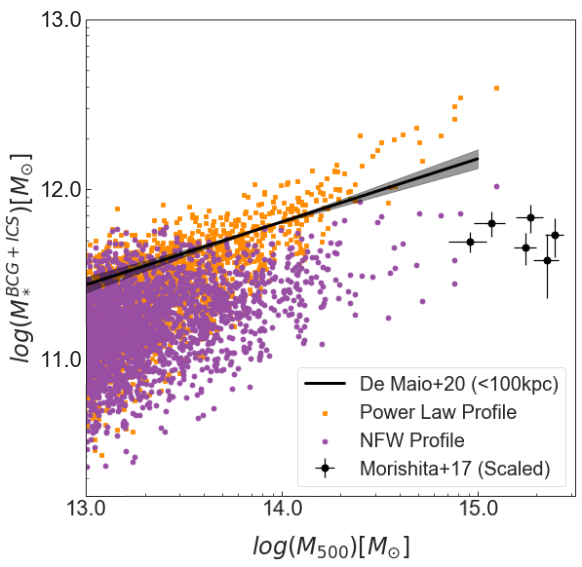}
    \caption{The total BCG+ICS stellar mass plotted with the results of \citet{2020MNRAS.491.3751D} and \citet{2017ApJ...846..139M}. The former is at a redshift of 0.4, and the latter at a mean redshift of 0.42. The data of \citet{2017ApJ...846..139M} has been from rescaled from its original aperture of 300kpc to 100kpc using an NFW profile, to allow for a fair comparison to \citet{2020MNRAS.491.3751D}. We include \lgaltt{}-MM ICS masses for a power-law profile (orange squares), and an NFW profile (purple points).}
    \label{fig:MBCGICSHT09}
\end{figure}

Fig. \ref{fig:MBCGICSHT09} shows the BCG+ICS mass for galaxy clusters as a function of $M_{500}$ for \lgaltt{}-MM and observations from \citet{2017ApJ...846..139M} and \citet{2020MNRAS.491.3751D}. All results are scaled to an aperture of 100 kpc here, to provide consistency among the data sets plotted. For \lgaltt{}-MM, we show this relation when assuming both a power-law stellar halo mass profile (orange points) and an NFW profile (purple points), with the gradual stripping model assumed in both cases.

\citet{2017ApJ...846..139M} do not fit their stellar density profiles with either a power-law or NFW profile, but instead perform third-order polynomial fits to the ICL.

\citet{2020MNRAS.491.3751D} study the evolution of the BCG+ICL by observing 42 groups and clusters. They find that, for a given $M\sub{500}$ mass, the stellar mass within 10-100 kpc increases across $\Bar{z}=1.55-0.40$, with there being minimal mass growth at lower redshifts, in contradiction to the conclusion drawn by \citet{2017ApJ...846..139M}. The authors infer that mass growth in the ICS occurs mainly within 100 kpc at early times, thereafter moving towards larger radii after $\Bar{z} = 0.4$. For Fig.~\ref{fig:MBCGICSHT09}, their intermediate redshift bin ($z=0.4$) was chosen, as this was closest to the redshift used by \citet{2017ApJ...846..139M}, namely $z=0.42$. \lgaltt{}-MM data at approximately $z=0.4$ is used as well.

When comparing the two observational data sets in Fig.~{\ref{fig:MBCGICSHT09}}, BCG+ICS masses are noticeably lower at fixed $M_{500}$ in the \citet{2017ApJ...846..139M} sample than in the \citet{2020MNRAS.491.3751D} sample (noting that the two samples only overlap by one system, MACS0416). This is most likely a result of the differences in mass calculation method. In particular, \citet{2017ApJ...846..139M} fit a third order polynomial to the mass surface density profile and integrate over it to obtain $M\sub{ICS}$. However, referring to their fig. 5, the two lowest-mass systems, Abell274 and MACS1149, only have ICL measurements out to approximately 175\,kpc. This makes inferences of the total ICS mass within 300\,kpc difficult to constrain from a polynomial fit.

The model results when considering an NFW profile show good agreement with the \citet{2017ApJ...846..139M} data. Taken in conjunction with Fig.~\ref{fig:MStellarHT09}, this suggests that the mass in satellite galaxies is over-estimated in \lgaltt{}-MM relative to these data, rather than the mass of the BCG itself.

When comparing to the \citet{2020MNRAS.491.3751D} data, we find that the \lgaltt{}-MM results when assuming a power-law profile show a slightly steeper slope. This is similar to the result found by \citet{2020MNRAS.491.3751D} for the \textsc{IllustrisTNG} simulation (their fig. 4), which exhibits power-law profiles in its model stellar haloes. When assuming an NFW profile, we instead find a similar slope but lower normalisation to that observed by \citet{2020MNRAS.491.3751D}. This emphasises the uncertainty associated with the choice of ICS profiles. The inherent non-uniformity of stellar halos and ICS components between systems means that these components cannot all be easily modelled by the same mass distribution.

With the inclusion of in-situ stars, the results of Fig. {\ref{fig:MBCGICSHT09}} would likely not change significantly. Only in-situ stars formed in the halo itself would contribute to an increase in $M_{*}^{\tn{BCG+ICS}}$, and such stars are believed to have a negligible contribution in cluster-sized systems \citep{Pillepich+18b}.

In conclusion, it appears that \lgaltt{}-MM over-predicts both ICS and total cluster stellar masses compared to the observations considered here. Although, this depends somewhat on the stripping model and stellar halo density profile assumed. Our analysis in this section suggests this result could be due to excessive star formation and stripping in satellite galaxies. However, over-production of stars in the progenitors of present-day BCGs could play a secondary role for the gradual stripping model. This is indicated in Fig.~{\ref{fig:SMF}}, which shows that the SMF for the gradual stripping model has a slight excess of very massive galaxies (as well as of L$^{*}$ galaxies) compared to the instantaneous stripping model. Apparent inconsistencies between the various observational data sets limit firmer conclusions at this time. In addition, we note that low-mass galaxies could be too faint to detect in massive clusters, leading to observational results under-estimating total cluster stellar masses compared to simulations.

\subsubsection{ICS mass fractions}\label{sec:ICSMassFractions}

\begin{figure*}
	\includegraphics[width=\textwidth]{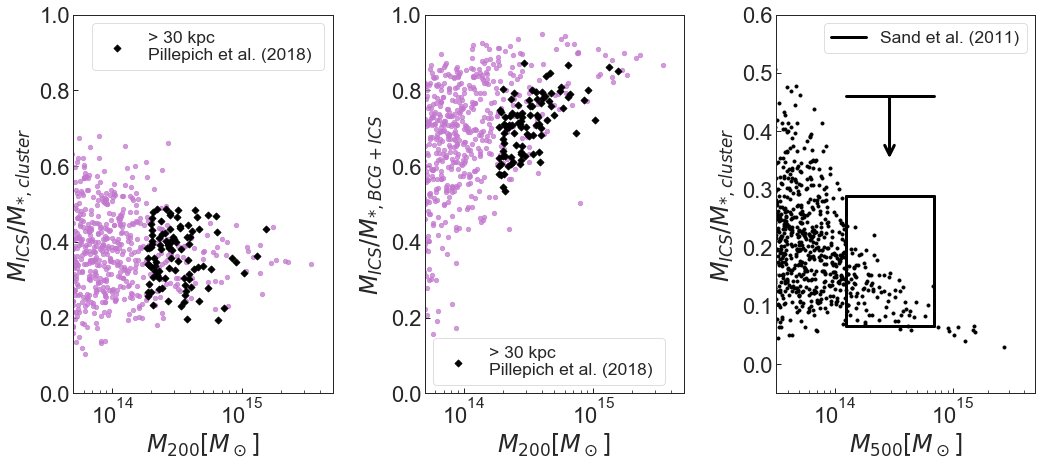}
    \caption{Two ICS mass fractions compared to the \textsc{IllustrisTNG} results of \citet{Pillepich+18b} (black diamonds). The left figure shows the ICS mass to total cluster stellar mass fraction, and the middle figure shows the ICS -- BCG+ICS stellar mass fraction. Both of these data sets are at redshift zero, measuring the ICS from 30\,kpc to the virial radius. In the rightmost figure, the ICS stellar mass fraction is plotted against $M\sub{500}$. The data of \citet{2011ApJ...729..142S} provides an upper limit, obtained when the authors assume stellar halo stars are exclusively old, and a boxed region, found when assuming stellar halo stars are of similar age to the stars in the central galaxy. Here, an aperture of $R_{200}$ is used at a redshift of 0.1.}
    \label{fig:ICSFracs}
\end{figure*}

In Fig. \ref{fig:ICSFracs}, the $M\sub{ICS}/M_{*\tn{,cluster}}$ (left panel) and $M\sub{ICS}/M_{*\tn{,BCG+ICS}}$ (middle panel) fractions for \lgaltt{}-MM (assuming the gradual stripping model) are compared to those from the \textsc{IllustrisTNG} simulation \citep{Pillepich+18b}. Those authors study the stellar mass in approximately 4000 model galaxy groups and clusters, focusing on high-mass systems. These results are at redshift zero, using an annular aperture between 30\,kpc and $R\sub{vir}$, which we also adopt for \lgaltt{}-MM here.

Our results agree very well with those from \textsc{IllustrisTNG}, suggesting that \lgaltt{}-MM is representing the apportionment of stellar mass among various components in galaxy clusters in a very similar way to magneto-hydrodynamical simulations. This is an encouraging sign, given the simpler modelling of stellar halo formation in semi-analytic models.

In the right-hand panel of Fig. \ref{fig:ICSFracs}, we show the observational results of \citet{2011ApJ...729..142S}, who obtain stellar halo masses within $R_{200}$ through the study of intracluster SNe-Ia at a mean redshift of $\sim0.1$. The authors provide two results for $M\sub{ICS}/M_{*\tn{,cluster}}$. The upper limit arises from the assumption that stellar halo stars are exclusively old. The box corresponds to the assumption that stellar halo stars have similar ages to galactic stars in the central galaxy. ICS mass fractions in \lgaltt{}-MM when assuming the gradual stripping model show very good agreement with this latter observational estimate. They also show good qualitative agreement with the observational findings of \citet{Burke+15,Jimenez-Teja&Dupke16,2017ApJ...846..139M}, which report ICL light fractions of $\lesssim{}20$ per cent for the most massive clusters at slightly higher redshifts ($0.2\lesssim{}z\lesssim{}0.4$). In contrast, the ICS fractions returned by the instantaneous stripping model are too low compared to these observations, especially in the galaxy group regime.

We note here that mean stellar halo ages are not explicitly tracked in \lgaltt{}. A comparison between the observed and modelled ages of this extragalactic stellar component will be the focus of future work.

With the inclusion of in-situ stars, the \lgaltt{}-MM data in the left panel of Fig. {\ref{fig:ICSFracs}} would likely move towards higher $M\sub{ICS}/M_{*\tn{,cluster}}$, thereby reducing agreement with \textsc{IllustrisTNG}. The middle panel would also see an increase, as heated stars from the BCG would raise the ICS mass by $\sim{}10-20$ per cent. As with the first panel, the right-most panel would see an increase in ICS fraction, but our model points would likely not shift outside the boxed region indicated by the observations of \citet{2011ApJ...729..142S}.

Finally, \citet{2009ApJ...691.1787S} measure ICS fractions within an observational aperture of $0.6\,R_{200}$ for twelve galaxy clusters with virial masses above $\sim1.6\times10^{14}\Msun$ at $z\lesssim0.1$. An equivalent population was selected in \lgaltt{}-MM (assuming the gradual stripping model), resulting in 149 galaxy clusters, and the properties of these clusters were averaged. The main results from this comparison are as follows:

\begin{itemize}
\item Using a simple chemical evolution model, \citet{2009ApJ...691.1787S} found that between $21\pm{}9$ and $31^{+11}_{-9}$ per cent of the total cluster ICM iron content is produced by the BCG+ICS, depending on assumptions about the SNe-Ia DTD. In \lgaltt{}-MM, this value was found to be only $6.2\pm0.8$ per cent. Of the remaining enrichment, $55\pm9$ was from satellite galaxy outflows and $30$ per cent was from cold and hot gas stripping of satellites.

\item \citet{2009ApJ...691.1787S} found that the ICS constitutes 81.04 per cent of the total ICM enrichment from the BCG+ICS. In \lgaltt{}-MM, this was found to be $88\pm3$ per cent. 


\item \citet{2009ApJ...691.1787S} found that the ICM iron content produced by the ICS alone was 25 per cent. In \lgaltt{}-MM, this was found to be $5.4\pm0.6$ per cent. 
\end{itemize}

The ICS/(BCG+ICS) iron enrichment fraction agrees well with the results of \citet{2009ApJ...691.1787S}. However, the BCG+ICS iron enrichment and the ICS iron enrichment show significant disagreement. The enrichment from stellar haloes is already fairly strong in \lgaltt{}-MM, with the iron abundances in MWA haloes being of an acceptable value (Figs. \ref{fig:FeHHT09} and \ref{fig:FeHHT09Scaled}), so it does not appear to be a case of stellar halo enrichment being too weak in our model. 

Another possibility is that ICS components do not provide enough enrichment because they are not massive enough. Our comparisons to observational data, however, suggest the masses in \lgaltt{}-MM are, if anything, over-predicted in clusters.

When checking these enrichment percentages for the instantaneous stripping model, there was a greater discrepancy with the \citet{2009ApJ...691.1787S} results. Namely, 2.8$\pm$0.7 per cent of the ICM iron content is produced by the BCG+ICS, the ICS produced 70$\pm$6 per cent of the total BCG+ICS iron enrichment, and the ICS produced 2.0$\pm$0.5 per cent of the total ICM iron content. These reductions in ICS enrichment are due to the reduced ICS masses in the instantaneous stripping model. 

Furthermore, we find that selecting different SNe-Ia DTD models does not improve agreement either. Both the Gaussian and bi-modal DTDs return worsened agreement with \citet{2009ApJ...691.1787S}, with all three fractions being of a lower value in comparison to the default power-law DTD.

Therefore, this discrepancy is likely an issue of differences in the chemical enrichment modelling between \lgaltt{}-MM and \citet{2009ApJ...691.1787S}. In particular, the modified version of \lgaltt{} used here assumes a SNe-Ia progenitor fraction of $A\sub{SNIa} = 0.035$, in line with many other works which constrain its value using Milky Way or early-type galaxy (ETG) observations \citep{Greggio+05,Matteucci+06,Calura&Menci09,Arrigoni+10a}. In contrast, \citet{2009ApJ...691.1787S} use a much higher fraction of 0.175. This value sets how many stars become SNe-Ia progenitors, and therefore affects ICM enrichment.

When assuming a value of $A\sub{SNIa} = 0.175$ in the \lgaltt{} model, the fractions found above increase, but not enough to reconcile the differences seen between \lgaltt{}-MM and \citet{2009ApJ...691.1787S}. These values become $7.6\pm0.9$ per cent, $91\pm3$ per cent, and $6.9\pm0.7$ per cent, respectively, for the gradual stripping model and assuming a power-law SN-Ia DTD.

Unfortunately, \citet{2009ApJ...691.1787S} does appear to be a fairly unique study, with there being little in the way of comparable papers which predict the enrichment from the ICS in comparison to other cluster stellar populations. However, a number of factors suggest that the \citet{2009ApJ...691.1787S} results may be over-estimates.

Firstly, the SNe-Ia rate of $0.066^{+0.027}_{-0.020}$ SNuM (SNe per unit mass) from \citet{2008MNRAS.383.1121M} that is used by \citet{2009ApJ...691.1787S} to tune their value of $A\sub{SNIa}$ is larger than the $0.041\pm0.018$ SNuM found by \citet{2012AAS...21910808S} using observations of galaxy clusters in the MENeaCS survey.

Secondly, their assumed value of $A\sub{SNIa} = 0.175$ is over twice as large as even the more liberal estimates used in the previous literature (see \eg{}\citealt{2004ApJ...604..579P,2012MNRAS.426.3282M}). For example, \citet{2013ApJ...773...52L} shows that standard assumptions on the IMF, initial metallicities, formation efficiencies, etc., result in an underprediction of ICM metallicities by a factor of at least two. The author describes a number of possible solutions through the use of modelling, with an increase to $A\sub{SNIa}$ being one (denoted as $\varepsilon^{\rm{Ia}}$ in their study). As a default value, \citet{2013ApJ...773...52L} adopts $A\sub{SNIa}=0.076$, but in an altered model considers a boosted value of $A\sub{SNIa}=0.13$, which is still lower than \citet{2009ApJ...691.1787S}'s assumption. Furthermore, while it does improve results, the author states that a physical motivation would need to be provided for such a large increase in progenitor fraction.

Thirdly, when assessing the \citet{2009ApJ...691.1787S} result that the ICS produces 25 per cent of the cluster ICM metallicity, \citet{2021arXiv210504638B} point out that, if this were the case, studies would show the ICS accounting for 80 per cent of the total cluster light, which is four times greater than what is currently observed. On this point, it is worth noting that \citet{2021arXiv210504638B} suggest the above result of \citet{2009ApJ...691.1787S} is four times too high, while we find a value for the same fraction around four to five times lower than \citet{2009ApJ...691.1787S}.

In combination, the above results suggest that the parameters chosen by \citet{2009ApJ...691.1787S} could lead to an over-estimation of the ICS's contribution to the enrichment of the ICM. Further observations, therefore, would be needed in order to confidently determine the accuracy of our results (see below).

We also note that, despite the discrepancies between the results of \citet{2009ApJ...691.1787S} and our gradual stripping results, \lgaltt{}-MM still manages to reproduce the total cluster iron abundances at low redshift well. This is demonstrated by \citet{Yates+21a}, who show that, when scaling and homogenising observational and model data in the same way (see \citealt{Yates+17}), the mass-weighted iron abundances in the hot gas surrounding galaxy groups and clusters (as measured by XMM-Newton and other X-ray telescopes) can be reproduced by \lgaltt{}-MM, simultaneously with the metallicities observed in the gas and stars within nearby star-forming galaxies. 

\subsubsection{ICS enrichment}\label{sec:BCGICSoutflows}

\begin{figure*}
	\includegraphics[width=.8\textwidth]{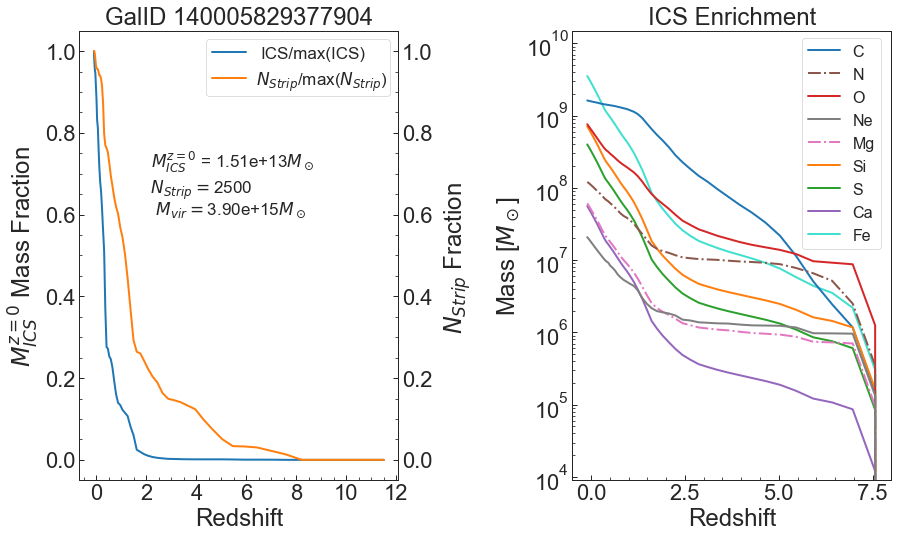}
    \caption{The evolution of the most massive galaxy in \lgaltt{}-MM, following the same analysis as the MWA studied in Fig. \ref{fig:MWANDis}. Left: The blue curve and orange curve describes the evolution of the ICS mass and stellar stripping count, respectively. Redshift zero ICS mass, stellar stripping number, and virial mass are annotated. Right: The cumulative ICM enrichment history of ICS stars in the cluster environment of this central galaxy. }
    \label{fig:Tree0Evo}
\end{figure*}

Lastly, in this section we present the stellar halo enrichment of the ICM for the most massive galaxy cluster in \lgaltt{}-MM, which has a virial mass of $M\sub{200}=3.90\times10^{15}\Msun$. This system has induced significantly more stripping events than the example MWA discussed in Section \ref{Disruption Events}. The left panel of Fig. \ref{fig:Tree0Evo} shows that the ICS of this massive cluster first forms at a redshift of $\sim8$, while the MWA's stellar halo first forms at redshift 4.6 (with the average for MWAs being $z=1.24$). By $z=0$, the ICS mass of this cluster is also four orders of magnitude greater than the MWA's stellar halo, and three orders of magnitude more massive than the average of the MWA sample ($4.02\times10^9\Msun$). As a result of this constant growth through stellar stripping events, the enrichment of the ICM by the ICS is essentially continuous (see the right panel of Fig. {\ref{fig:Tree0Evo}}).

\begin{figure*}
	\includegraphics[width=.9\textwidth]{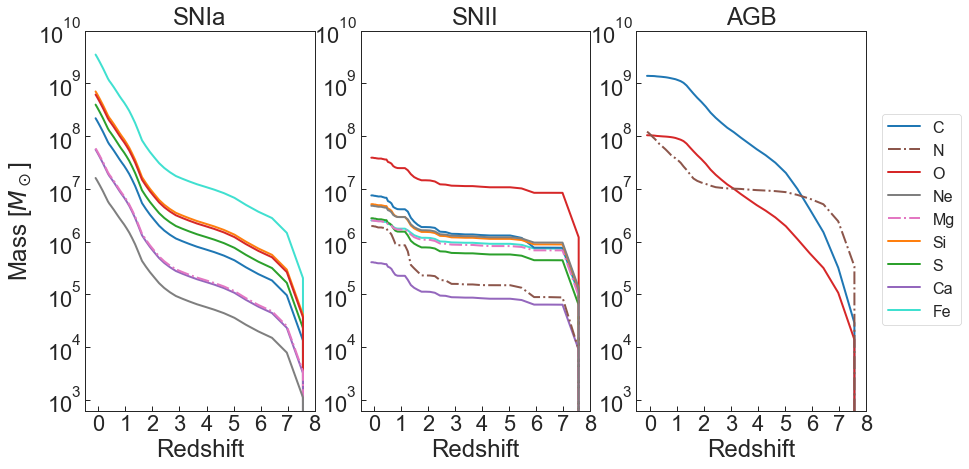}
    \caption{The cumulative ICM enrichment of Fig. \ref{fig:Tree0Evo} split into individual sources. As for Fig. \ref{fig:MWASplit}, only the newly synthesised metals from SNe-Ia, SNe-II, or AGB stars are plotted here. Metals which pass through the progenitor stars unprocessed are not included.}
    \label{fig:Tree0EvoSplit}
\end{figure*}

\begin{figure*}
	\includegraphics[width=.8\textwidth]{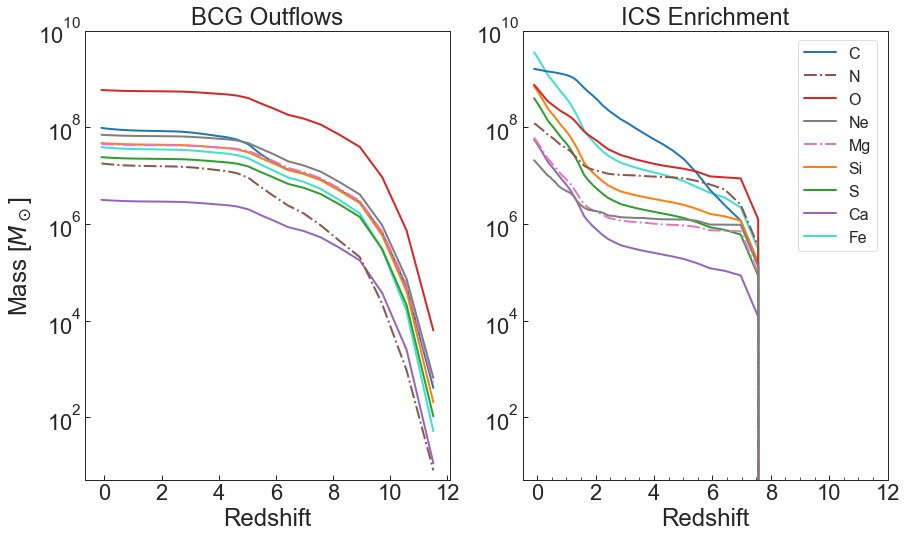}
    \caption{A comparison between the cumulative ICM enrichment from BCG outflows produced by this central galaxy and from the ICS of the surrounding cluster environment, following the same analysis as the MWA in Fig. \ref{fig:MWAOutflows}. }
    \label{fig:BCGOutflow}
\end{figure*}

When comparing enrichment from individual enrichment channels in MWAs and galaxy clusters (\ie{}comparing Fig.~{\ref{fig:MWASplit}} with Fig.~{\ref{fig:Tree0EvoSplit}}), we can see that enrichment from SNe-Ia is more significant in clusters at low redshift, due to the significant accretion of (older) stellar populations into the ICS component at late times (see also \citealt{Burke+15}). SNe-II in the ICS show similar enrichment characteristics to those in MWA stellar haloes, but only up to $z\approx2$. Thereafter, the carbon mass ejected becomes significantly larger -- a known feature of the SNe-II yields used in \lgaltt{} for high-mass, metal-rich stars (see fig. 4 in \citealt{2013MNRAS.435.3500Y}). For AGB stars, both the MWA shown in Fig.~{\ref{fig:MWASplit}} and the massive cluster shown in Fig. {\ref{fig:Tree0EvoSplit}} exhibit around 10 times greater C production than O production by $z=0$. However, overall, the mass difference between C and O is lower in the cluster galaxy than the MWA (right-hand plots of Fig. \ref{fig:MWANDis} and \ref{fig:Tree0Evo}), likely a result of the larger oxygen contribution from low-redshift SNe-Ia (Fig. \ref{fig:Tree0EvoSplit}).

Fig. {\ref{fig:BCGOutflow}} shows that, by redshift 0, the ICS of this massive cluster has enriched the surrounding ICM more significantly than outflows from the BCG have, for most elements. This is in stark contrast to the case for MWAs, where stellar halo enrichment is around two orders of magnitude lower than that from outflows. The ICS enrichment of carbon, for example, is over an order of magnitude greater than what is produced by outflows from the BCG. Around $z=4$, star formation in the BCG tails off significantly as gas is depleted, resulting in reduced ICM enrichment. Constant stellar stripping of satellite galaxies into the ICS, however, produces a steady supply of stars, and, as a result, enrichment.

To conclude this section, we find interesting differences between the enrichment of the hot CGM/ICM by stellar halo stars in MWAs and massive galaxy clusters. The stellar haloes of massive galaxy clusters (\ie{}the ICS) play a much more significant role in enrichment, particularly at late times, even exceeding the cumulative enrichment provided by outflows from the BCG. The chemical composition of the enriching material from halo stars is also different in MWAs and galaxy clusters. The ICS ejects a much higher fraction of iron and heavier alpha elements into the surrounding hot gas by $z=0$, mainly due to a high rate of SNe-Ia at late times. Instruments on board future X-ray observatories such as \textit{XRISM/Resolve} \citep{Tashiro+20} and \textit{Athena/X-IFU} \citep{Barret+18} should be able to probe such chemical signatures in the hot gas surrounding galaxy groups and clusters to high precision, thus helping to confirm or deny the predictions provided by \lgaltt{}-MM here.

%% file: Chapters/Conclusions.tex
This paper presents the formation and chemical enrichment of stellar haloes in the \lgaltt{} semi-analytic model using a modified gradual stripping model for stars and cold gas in satellite galaxies. The modified version of the model was chosen for our analysis, \lgaltt{}-MM, which allows up to 90 per cent of the material ejected by SNe in galaxy discs to directly enrich the hot gas surrounding galaxies. Following methods similar to the literature, good agreement is obtained across a range of results for MWA systems and galaxy clusters in \lgaltt{}-MM. Our main results are summarised below.

\begin{itemize}
    \item The gradual stripping model implemented into \lgaltt{}-MM results in an increased number of L* galaxies, in comparison to the instantaneous stripping model (see Section \ref{sec:TidalDisruptions}). The greater stripping efficiency in the gradual stripping model leads to less gas being accreted onto central galaxies and their supermassive black holes, thereby reducing the AGN feedback efficiency and increasing star formation in these systems.
    
    \item \lgaltt{}-MM is able to reasonably reproduce the slope of the observed stellar halo mass -- metallicity relation when a simple linear metallicity profile is applied and [Fe/H] is measured at the same aperture as the literature. However, the normalisation of this relation is higher in the simulation than observed in local MWAs (see Section \ref{sec:FeH}). The inclusion of an in-situ stellar halo formation mechanism into \lgaltt{}-MM may help improve the normalisation of the stellar halo mass -- metallicity relation.
    
    \item At the initial formation of the stellar halo, the CGM surrounding MWAs in \lgaltt{}-MM is predominantly enriched with oxygen by SNe-II whose progenitors were stripped from satellite galaxies shortly after star formation. Carbon then becomes the dominant element ejected by stellar halo stars into the CGM, a result of winds from AGB stars. At late-times, iron from SNe-Ia dominates, particularly if a Gaussian SNe-Ia DTD with no significant prompt or delayed component is assumed. On average, the stellar halo produces only 0.1 per cent of the total CGM enrichment in MWAs, and satellite galaxies only contribute a further 0.05 per cent. The stellar haloes in MWA galaxies have a mean $z=0$ mass of $4.02\times10^9\Msun$ in \lgaltt{}-MM.
    
    \item Similarly to MWAs, the ICS in galaxy clusters contributes mainly oxygen at early times and carbon at intermediate times, with the iron enrichment from SNe-Ia being strongest at late times. However in contrast, ICS components can exceed the enrichment from the central BCG (see Section \ref{sec:BCGICSoutflows}). The ICS produces only 5.40 per cent of the total ICM iron enrichment in massive clusters by $z=0$, with the majority of the enrichment coming from satellite galaxies.
    
    \item For our gradual stripping model, 55 per cent of MWAs had stellar haloes formed from only one or two significant progenitors. For those with one significant progenitor, a mean mass of $2.33\times10^9\Msun$ was contributed, similar to the mass measured for Gaia Enceladus in the Milky Way's inner stellar halo. MWA stellar halo masses in \lgaltt{}-MM are slightly lower than those seen in the \textsc{Auriga} simulation, even when only considering ex-situ stars in both simulations. However, \lgaltt{}-MM is in good agreement with estimates of the real Milky Way stellar halo mass. 
    
    \item ICS mass fractions in \lgaltt{}-MM agree well with observations when using the gradual stripping model in combination with an NFW mass profile (see Section \ref{sec:ICSMasses}). Good correspondence to the ICS mass fractions in \textsc{TNG300} was also found. However, total stellar masses, BCG+ICS masses, and ICS masses alone were consistently over-estimated in our model. \lgaltt{}-MM, therefore, likely requires a reduction in the overall cluster stellar mass formation in order to improve agreement with observations.

\end{itemize}

The implementation of more realistic and self-consistent mass, density, and metallicity profiles for stellar haloes in \lgaltt{} is one promising avenue for further study. Likewise, the inclusion of an in-situ stellar halo formation mechanism would allow a fairer and more complete comparison with observational data on the metal-richer inner halo of the Milky Way. The inclusion of key stellar halo properties into the MCMC formalism used to constrain free parameters in \lgaltt{} will also be considered in future work. Lastly, calculating the ages and [$\alpha$/Fe] ratios of stellar haloes in \lgaltt{} would open-up a new dimension of comparisons with MWA and galaxy cluster data, helping to further constrain the formation and chemical evolution of stellar haloes across cosmic time.